\documentclass{ws-ijmpe}
\usepackage[super,compress]{cite}
\usepackage{multirow}
\usepackage{float}
\usepackage[colorlinks=true,citecolor=blue,urlcolor=blue,linkcolor=red]{hyperref}
\begin{document}

\catchline{}{}{}{}{}

\title{BigApple force and its implications to finite nuclei and astrophysical objects}

\author{
H. C. Das$^{1,2}$\footnote{harish.d@iopb.res.in},
Ankit Kumar $^{1,2}$,
Bharat Kumar $^{3}$,
S. K. Biswal$^{4}$,
S. K. Patra$^{1,2}$}

\address{$^{1}$ Institute of Physics, Sachivalaya Marg, Bhubaneswar 751005, India
}
\address{$^{2}$ Homi Bhabha National Institute, Training School Complex, Anushakti Nagar, Mumbai 400094, India}
\address{$^{3}$ Department of Physics $\&$ Astronomy, National Institute of Technology, Rourkela 769008, India }
\address{$^{4}$ Department of Engineering Physics, DRIEMS Autonomous Engineering College, Cuttack 754022, India}

\maketitle

\begin{history}
\end{history}

\begin{abstract}
The secondary component of the GW190814 event left us with a question, ``whether it is a supermassive neutron star or lightest black-hole?". Recently, Fattoyev {\it et al.} have obtained an energy density functional (EDF) named as BigApple, which reproduces the mass of the neutron star is 2.60 $M_\odot$ which is well consistent with GW190814 data. This study explores the properties of finite nuclei, nuclear matter, and neutron stars by using the BigApple EDF along with four well-known relativistic mean-field forces, namely NL3, G3, IOPB-I, and FSUGarnet. The finite nuclei properties like binding energy per particle, skin thickness, charge radius, single-particle energy, and two-neutron separation energy are well predicted by the BigApple for a series of nuclei. The calculated nuclear matter quantities such as incompressibility, symmetry energy, and slope parameters at saturation density are consistent with the empirical or experimental values where ever available. The predicted canonical tidal deformability by the BigApple parameter set is well-matched with the GW190814 data. Also, the dimensionless moment of inertia lies in the range given by the analysis of PSR J0737-3039A.
\end{abstract}


\ccode{PACS numbers: 21.10.Dr; 21.10.Pc; 21.65.+f;  26.60.+c; 97.60.Jd; 04.30.-w}

\section{Introduction}
\label{intro}
The detection of gravitational wave (GW) from the merger of two neutron stars (NSs) by the LIGO and Virgo collaboration (LVC) provides us with an opportunity to study the properties of dense matter at extreme conditions \cite{Abbott_2017, Abbott_2018}. Further, with the recent discovery event GW190814, the LVC detected the collision of a black hole of mass 22.2--24.3 $M_\odot$ with a compact object of mass 2.50--2.67 $M_\odot$ \cite{RAbbott_2020}. There is no measurable signature of tidal deformability and electromagnetic counterpart on the gravitational waveform, which left us in doubt, and raises the question, ``whether the secondary component of GW190814 is either a massive NS or lightest black hole?".

Recently some works have been done to explore the secondary component of GW190814 \cite{Most_2020,Vattis_2020,Tews_2021,Roupas_2021,Godzieba_2021,Huang_2020}. Tan \textit{et al.} \cite{Tan_2020} considered that it may be a heavy NS with deconfined QCD core. In the Ref. \cite{ZhangAAS_2020}, they assumed that it might be a super-fast pulsar. Recently, Fattoyev \textit{et al.} \cite{Fattoyev_2020} have assumed it to be a binary black-hole merger, which is supported by Tews \textit{et al.} \cite{Tews_2021}. Using the DDRMF model in Ref. \cite{Huang_2020}, it is reported that there is a possibility of the secondary component in GW190814 as a  NS. Recently Das {\it et al.} reported that the secondary component might be the dark matter admixed NS with EOS is sufficiently stiff \cite{DasPRD_2021}. Therefore, the GW190814 event opens the possibility to achieve the supermassive NS only (i) when the equation of state (EOS) is very stiff, or (ii) it is rapidly rotating with mass shedding frequency \cite{ZhangAAS_2020, Biswas_2021}. Thus, one can consider either of the two facts, which can reproduce the mass of the secondary component of GW190814 around 2.50 $M_\odot$.
  	
The binary NS merger event GW170817 \cite{Abbott_2017} and its electromagnetic counterpart \cite{Margalit_2017} has evoked the astrophysics community to put an upper bound on the maximum mass of the non-rotating NS. By combining the total binary mass of GW170817 inferred from GW signal with electromagnetic observations, an upper limit of $M_{max}$ $\leq$  2.17 $M_\odot$ is predicted by Margalit {\it et al.} \cite{Margalit_2017}. Rezzolla {\it et al.} have put the upper bound by combing the GW observations and quasi-universal relations with $M_{max}\leq 2.16_{-0.15}^{+0.17} \ M_\odot$ \cite{Rezzolla_2018}. Further analysis employing both energy and momentum conservation laws and the numerical-relativity simulations show that the maximum mass of a cold NS is constrained to be $M_{max}\leq 2.3 \ M_\odot$ \cite{Shibata_2019}. Also, different massive pulsar discoveries constrained the EOS of the supra-nuclear matter inside the core of the NS \cite{Demorest_2010, Antoniadis_2013, Cromartie_2019}. These observational data also put strong constraints on the maximum mass of the slowly rotating NS with a lower bound $\sim$ 2 $M_\odot$, which discarded many EOSs. Recently, the Neutron Star Interior Composition Explorer (NICER) data also put stringent constraints on both mass and radius of a canonical star from the analysis of PSR J0030+0451 \cite{Miller_2019, Riley_2019, Raaijmakers_2019}. However, one can not exclude the existence of a supermassive NS as the secondary component of GW190814. 
	
Many energy density functionals (EDFs) have been constructed by using various effective theory approaches, like relativistic mean-field (RMF) formalism  \cite{Lalazissis_1997,Frun_1997,Reinhard_1988,Singh_2013,Kumar_2017,Kumar_2018}, Skyrme-Hartree-Fock (SHF) \cite{Chabanta_1998,Brown_1998,Stone_2007,Dutra_2012}, density-dependent RMF (DDRMF) \cite{Niksic_2002,Typel_2005,Lalazissis_2005,Klahn_2006} etc. to study the finite nuclei, the nuclear matter and the NS properties. But only a few parameter sets are able to reproduce the properties consistently with empirical/experimental data \cite{Dutra_2012, Dutra_2014}. To achieve the mass of the secondary component for the GW190814 event, one has to take very  stiff EOS, which may or may not satisfy different constraints such as flow data \cite{Danielewicz_2002}, GW170817 \cite{Abbott_2017, Abbott_2018}, X-ray \cite{Antoniadis_2013, Cromartie_2019}, and NICER data \cite{Miller_2019, Riley_2019}.
It is not a big deal to form an EDF that predicts supermassive NS with a mass range of 2.50-2.67 $M_\odot$ but to reconcile the supermassive NS with different properties finite and infinite nuclear systems at the same time are quite difficult. The GW170817 has put a limit on the tidal deformability of the canonical NS, which is used to constraint the EOS of neutron-rich matter at 2--3 times of the nuclear saturation densities \cite{Abbott_2017}. Due to a strong dependence of the tidal deformability with radius ($\Lambda \sim R^5$), it can put stringent constraints on the EOS. Several approaches \cite{Bauswein_2017,Annala_2018,Fattoyev_2018,Radice_2018,Mallik_2018,Most_2018,Tews_2018,Nandi_2019,Capano_2020} have been tried to constraint the EOS on the basis of the tidal deformability bound given by the GW170817. In particular, the GW170817 favors a star with a relatively smaller radius, which needs a softer EOS. These constraints are also consistent with the recent observations of mass-radius by the NICER with the X-ray study of the millisecond pulsar PSR J0030+0451. \cite{Miller_2019,Riley_2019,Bogdanov_2019,Raaijmakers_2019}. 

There is a tension for reconciliation between the two facts (i) prediction of supermassive NS, which requires stiffer EOSs, and (ii) softer EOSs are required to satisfy the observation data GW170817 and NICER. Fattoyev \textit{et al.} have been tried to reconcile the two facts, which are based on the density functional theory. They have predicted the NS maximum mass as $2.60 \ M_\odot$ from the covariant EDF, named `BigApple'  \cite{Fattoyev_2020}. Here, we briefly describe the formalism of the BigApple EDF. Fattoyev \textit{et al.} have tuned the coupling constant for $\omega^4$ interaction  ($\zeta_0$) in such a way that, which predicts the mass of the NS found to be 2.60 $M_\odot$. To reproduce the other NM properties, they have slightly changed the symmetry energy slope, which predicts the skin thickness of $^{208}$Pb as 0.15 fm. The binding energies and charge radii of $^{40}$Ca and $^{208}$Pb have been fixed by the re-adjusting the mass of the $\sigma$-meson, the NM saturation density, and the binding energy at the saturation \cite{Fattoyev_2020}. 

The main aim for the construction of the BigApple parameter set is to study the secondary component of the GW190814, i.e., basically the properties of a supermassive NS. But in this case, we extensively use this parameter set to study the different properties of finite nuclei, NM, and the NS. Fattoyev {\it et al.} have calculated the finite nuclei properties using the BigApple parameter set only for few nuclei, but in this present calculation, we take into consideration the superheavy and also a series of nuclei that extends from the proton to the neutron drip line. The properties of finite nuclei are calculated, for example, skin thickness, single-particle energy, two neutron separation energy, and isotopic shift using the BigApple EDF. After that, we extended our calculations to unfold the properties of NM and the mysteries of NS. Therefore, we checked the BigApple EDF in different systems to calibrate its validity and proper parameter set in these calculations.

The BigApple parameter set predicts not only the finite nuclei properties like binding energy per nucleon, charge radius, and skin thickness for a series of nuclei but also the some NM properties at the saturation density as well. The EOS of BigApple doesn't satisfy the heavy-ion data. The predicted dimensionless tidal deformability corresponding to the BigApple set isn't consistent with GW170817 due to its stiffer EOS. The calculated canonical radius lies in the range given by the NICER data. Therefore, Fattoyev {\it et al.} have concluded that the secondary component is the lightest black hole.

The paper is organised as follows: the formalism used in this work is given in Section \ref{formalism}. In Sub-Section \ref{EDF}, we describe the basic formalism of the RMF approach. In Sub-Section \ref{FormNM}, we calculate the NM properties such as EOS, incompressibility, symmetry energy and its different coefficients, etc., for five known parameter sets. We present the results and discussion in Section \ref{r&d}, which includes finite nuclei, NM, and NS calculation in detail. Finally, we give concluding remarks in Section \ref{summary}.

\section{Formalism}
\label{formalism}
\subsection{Energy density functional}
\label{EDF}
In RMF prescription, the nucleons interact by exchanging different mesons like $\sigma$, $\omega$, $\rho$, and $\delta$. We take a model which includes all possible interactions for nucleon-mesons and their self and cross interactions up to fourth-order (except $\rho^4$) known as the effective field theory motivated relativistic mean-field (E-RMF). The parameters used in the E-RMF formalism are fitted to reproduce the observables of finite nuclei and infinite NM. The E-RMF EDF is given in Refs. \cite{Frun_1997,Singh_2014,Kumar_2017,Kumar_2018} as follow
\begin{eqnarray}
{\cal E}({r})&=&\sum_{\alpha=p,n} \varphi_\alpha^\dagger({r})\Bigg\{-i \mbox{\boldmath$\alpha$} \!\cdot\!\mbox{\boldmath$\nabla$}+\beta \bigg[M-\Phi (r)-\tau_3 D(r)\bigg]W({r})+\frac{1}{2}\tau_3 R({r})+\frac{1+\tau_3}{2}A({r})
\nonumber\\
&-&\frac{i\beta\mbox{\boldmath$\alpha$}}{2M}\!\cdot\!\bigg(f_\omega\mbox{\boldmath$\nabla$}W({r})+\frac{1}{2}f_\rho\tau_3 \mbox{\boldmath$\nabla$}R({r})\bigg)\Bigg\} \varphi_\alpha(r)-\frac{\zeta_0}{4!}\frac{1}{g_\omega^2 }W^4({r})+\Big(\frac{1}{2}+\frac{\kappa_3}{3!}\frac{\Phi({r})}{M}
\nonumber\\
&+&\frac{\kappa_4}{4!}\frac{\Phi^2({r})}{M^2}\Big)
\frac{m_s^2}{g_s^2}\Phi^2({r})+\frac{1}{2g_s^2}\times\left(1+\alpha_1\frac{\Phi({r})}{M}\right)\bigg(
\mbox{\boldmath $\nabla$}\Phi({r})\bigg)^2-\frac{1}{2g_\omega^2}\left( 1 +\alpha_2\frac{\Phi({r})}{M}\right)
\nonumber\\
&&\bigg(\mbox{\boldmath$\nabla$} W({r})\bigg)^2-\frac{1}{2}\Big(1+\eta_1\frac{\Phi({r})}{M}+\frac{\eta_2}{2}\frac{\Phi^2({r})}{M^2}\Big)\frac{m_\omega^2}{g_\omega^2} W^2({r})-\frac{1}{2e^2} \bigg( \mbox{\boldmath $\nabla$} A({r})\bigg)^2
\nonumber\\
&-&\frac{1}{2g_\rho^2} \bigg( \mbox{\boldmath $\nabla$} R({r})\bigg)^2 -\frac{1}{2} \left( 1 + \eta_\rho \frac{\Phi({r})}{M}\right)\frac{m_\rho^2}{g_\rho^2} R^2({r})-\Lambda_{\omega}\bigg(R^{2}(r)\times W^{2}(r)\bigg)
\nonumber\\
&+&\frac{1}{2 g_{\delta}^{2}}\left( \mbox{\boldmath $\nabla$} D({r})\right)^2+\frac{1}{2}\frac{{m_{\delta}}^2}{g_{\delta}^{2}}D^{2}(r)\;,
\label{lagran}
\end{eqnarray}
where $\Phi$, $W$, $R$ and $D$ are the redefined fields $\Phi = g_s\sigma$, $W = g_\omega \omega$, $R$ = g$_\rho\vec{\rho}$ and $D=g_\delta\delta$. The coupling constants $g_\sigma$, $g_\omega$, $g_\rho$, $g_\delta$  and the masses $m_\sigma$, $m_\omega$, $m_\rho$ and $m_\delta$ are respectively taken for $\sigma$, $\omega$, $\rho$, and  $\delta$ mesons. $\frac{e^2}{4\pi}$ is the photon coupling constant. The coupling constants and masses for different mesons are listed in Table \ref{table1} for five different parameter sets NL3 \cite{Lalazissis_1997}, FSUGarnet \cite{Chen_2015}, G3 \cite{Kumar_2017}, IOPB-I \cite{Kumar_2018} and BigApple \cite{Fattoyev_2020}. 

From Eq. (\ref{lagran}), we get the equation of motions for the mesons and nucleons  using the equation $\big(\partial\mathcal{E}/\partial\phi_i\big)$=0 at constant density. The field equations are solved self-consistently to get all the mesons fields. The mean-field equations for different mesons such as iso-scalar-scalar $\sigma$, iso-scalar-vector $\omega$, iso-vector-vector $\rho$, iso-vector-scalar $\delta$ are given in Refs. \cite{Kumar_2017, Kumar_2018}. The ground-state properties of finite and infinite nuclear systems are obtained numerically in a self-consistent iterative method. The total binding energy of the nucleus is written as 
\begin{eqnarray}
E_{total}= E_{part}+E_{\sigma}+E_{\omega}+E_{\rho}+E_
{\delta}+E_{c}+E_{pair}+E_{c.m.},
\end{eqnarray}
where $E_{part}$ is the sum of the single-particle energies of the nucleons and $E_{\sigma}$, $E_{\omega}$, $E_{\rho}$, $E_{\delta}$ are the energies of the respective mesons. $E_c$ is the energy from the Coulomb repulsion due to protons. $E_{c.m.} (=\frac{3}{4}41A^{-1/3}$ MeV) is the centre of mass energy correction evaluated with a non-relativistic approximation \cite{Negele_1970,MCentelles_2001}. The pairing energy $E_{pair}$ is calculated by assuming few quasi-particle level as discussed in Refs. \cite{DelEstal_2001,Kumar_2018}.
\subsection{Nuclear Matter Properties}
\label{FormNM}
\subsubsection{The equation of state of the NM:-}
To calculate the EOS for a NM system, one has to switched-off the Coulomb part. The energy density and pressure for the NM system are calculated as \cite{Kumar_2018, Das_2020}
\begin{eqnarray}
    \cal{E}&=& \frac{\gamma}{(2\pi)^{3}}\sum_{\alpha=p,n}\int_0^{k_\alpha} d^{3}k E_{\alpha}^\star (k_\alpha)+\rho_bW +\frac{1}{2}\rho_{3}R
    +\frac{ m_{s}^2\Phi^{2}}{g_{s}^2}\Bigg(\frac{1}{2}+\frac{\kappa_{3}}{3!}
    \frac{\Phi }{M} + \frac{\kappa_4}{4!}\frac{\Phi^2}{M^2}\Bigg)
    \nonumber\\
    &-&\frac{1}{4!}\frac{\zeta_{0}W^{4}}
    {g_{\omega}^2}-\frac{1}{2}m_{\omega}^2\frac{W^{2}}{g_{\omega}^2}\Bigg(1+\eta_{1}\frac{\Phi}{M}+\frac{\eta_{2}}{2}\frac{\Phi ^2}{M^2}\Bigg)-\Lambda_{\omega}(R^{2}\times W^{2})
    -\frac{1}{2}\Bigg(1+\frac{\eta_{\rho}\Phi}{M}\Bigg)\frac{m_{\rho}^2}{g_{\rho}^2}R^{2}
    \nonumber\\
    &+&\frac{1}{2}\frac{m_{\delta}^2}{g_{\delta}^{2}}D^{2},
    \label{enm}
    \end{eqnarray}
    and
    \begin{eqnarray}
    P&=&  \frac{\gamma}{3 (2\pi)^{3}}\sum_{\alpha=p,n}\int_0^{k_\alpha} d^{3}k \frac{k^2}{E_{\alpha}^\star (k_\alpha)}+\frac{1}{4!}\frac{\zeta_{0}W^{4}}{g_{\omega}^2}-\frac{ m_{s}^2\Phi^{2}}{g_{s}^2}\Bigg(\frac{1}{2}+\frac{\kappa_{3}}{3!}
    \frac{\Phi }{M}+ \frac{\kappa_4}{4!}\frac{\Phi^2}{M^2}\Bigg)
    \nonumber\\
    &+&\Lambda_{\omega} (R^{2}\times W^{2})+\frac{1}{2}m_{\omega}^2\frac{W^{2}}{g_{\omega}^2}\Bigg(1+\eta_{1}\frac{\Phi}{M}+\frac{\eta_{2}}{2}\frac{\Phi ^2}{M^2}\Bigg)+\frac{1}{2}\Bigg(1+\frac{\eta_{\rho}\Phi}{M}\Bigg)\times\frac{m_{\rho}^2}{g_{\rho}^2}R^{2}
    \nonumber\\
    &&
     -\frac{1}{2}\frac{m_{\delta}^2}{g_{\delta}^{2}}D^{2}.
    \label{pnm}
    \end{eqnarray}
    %
The $E_{\alpha}^\star(k_\alpha)$=$\sqrt {k_\alpha^2+{M_{\alpha}^\star}^2}$, where $k_\alpha$ is the momentum and $\gamma$ is the spin degeneracy factor which is equal to 2 for individual nucleons. The $M_\alpha^{\star}$ is the effective masses of nucleon given as
    \begin{eqnarray}
    M_{p,n}^\star &=& M + \Phi \mp D.
    \label{effnm}
    \end{eqnarray}
\begin{table} 
\centering
\caption{The NM parameters for NL3 \cite{Lalazissis_1997}, FSUGarnet \cite{Chen_2015}, G3 \cite{Kumar_2017}, IOPB-I \cite{Kumar_2018} and BigApple \cite{Fattoyev_2020} are listed. The mass of nucleon $M$ is 939 MeV and other coupling constants are dimensionless.}
\begin{tabular}{ccccccccccc}
\hline
\hline
\multicolumn{1}{c}{Parameter}
&\multicolumn{1}{c}{NL3}
&\multicolumn{1}{c}{FSUGarnet}
&\multicolumn{1}{c}{G3}
&\multicolumn{1}{c}{IOPB-I}
&\multicolumn{1}{c}{BigApple}\\
\hline
$m_{s}/M$  &  0.541  &  0.529&  0.559&0.533& 0.525 \\
$m_{\omega}/M$& 0.833  & 0.833 &  0.832&0.833& 0.833 \\
$m_{\rho}/M$&  0.812 & 0.812 &  0.820&0.812& 0.812 \\
$m_{\delta}/M$ & 0.0  &  0.0&   1.043&0.0&0.0  \\
$g_{s}/4 \pi$  &  0.813  &  0.837 &  0.782 &0.827&0.769 \\
$g_{\omega}/4\pi$ & 1.024 & 1.091 &  0.923&1.062&0.980 \\
$g_{\rho}/4 \pi$&  0.712  & 1.105&  0.962 &0.885&1.126 \\
$g_{\delta}/4 \pi$  &  0.0  &  0.0&  0.160& 0.0&0.0 \\
$k_{3} $   &  1.465  & 1.368&    2.606 &1.496&1.878 \\
$k_{4}$  &  -5.688  &  -1.397& 1.694 &-2.932&-7.382  \\
$\zeta_{0}$  &  0.0  &4.410&  1.010  &3.103&0.106 \\
$\eta_{1}$  &  0.0  & 0.0&  0.424 &0.0& 0.0 \\
$\eta_{2}$  &  0.0  & 0.0&  0.114 &0.0&  0.0\\
$\eta_{\rho}$  &  0.0  & 0.0&  0.645& 0.0 &0.0 \\
$\Lambda_{\omega}$& 0.0  &0.043 &  0.038&0.024& 0.047 \\
$\alpha_{1}$  &  0.0  & 0.0&   2.000&0.0&0.0 \\
$\alpha_{2}$  &  0.0  & 0.0&  -1.468&0.0 &0.0 \\
$f_\omega/4$  &  0.0  & 0.0&  0.220&0.0& 0.0\\
$f_\rho/4$  &  0.0  & 0.0&    1.239&0.0& 0.0\\
$\beta_\sigma$  &  0.0  & 0.0& -0.087& 0.0& 0.0 \\
$\beta_\omega$  &  0.0  & 0.0& -0.484& 0.0&0.0  \\
\hline
\hline
\end{tabular}
\label{table1}
\end{table}
\subsubsection{Symmetry energy and its different co-efficients:-}
The energy density ${\cal{E}}$ can be expanded in a Taylor series in terms of asymmetry factor $\xi  \big(=\frac{\rho_n-\rho_p}{\rho_n+\rho_p}\big)$ \cite{Horowitz_2014,Baldo_2016,Kumar_2018}.
    \begin{equation}
    {\cal{E}}(\rho,\xi) = {\cal{E}}(\rho,\xi=0)+S(\rho) \xi^2+ {\cal{O}}(\xi^4),
    \label{dere}
    \end{equation}
    where ${\cal{E}}(\rho,\xi = 0)$ is the energy of symmetric NM, $\rho$ is the baryonic density and $S(\rho)$ is the density dependence symmetry energy, which is defined as 
    \begin{equation}
    S(\rho) = \frac{1}{2}\Bigg(\frac{\partial^2{\cal{E}}}{\partial\xi^2}\Bigg)_{\xi=0}.
    \label{sym1}
    \end{equation}
    The nature of $S(\rho)$ is the most uncertain property of the NM. A lot of progress had been made both experimentally and theoretically to constrain the $S(\rho)$ in different density ranges, starting from heavy-ion collision to NS \cite{BaoLi_2013, Danielewicz_2002}. It has a large diversion at a high-density limit \cite{BaoLi_2019}. Here, we can expand the $S(\rho)$ in a leptodermous expansion near the saturation density. The expression  of density dependence symmetry energy is as follow \cite{Matsui_1981,Kubis_1997,MCentelles_2001,Chen_2014,Kumar_2018}:
    \begin{equation}
    S(\rho) = J+L\eta+\frac{1}{2}K_{sym}\eta^2+\frac{1}{6}Q_{sym}\eta^3+{\cal{O}}(\eta^4),
    \label{eq11}
    \end{equation}
    where $\eta$=$\frac{\rho-\rho_0}{3\rho_0}$, $J$ is the symmetry energy at saturation density $\rho_0$ and the other parameters like slope ($L$), curvature ($K_{sym}$) and skewness ($Q_{sym}$) are given as follow:
    \begin{eqnarray}
    L=3\rho\frac{\partial S(\rho)}{\partial\rho}\Big|_{\rho=\rho_0},\\
    K_{sym}=9\rho^2\frac{\partial^2 S(\rho)}{\partial\rho^2}\Big|_{\rho=\rho_0},\\
    Q_{sym}=27\rho^3\frac{\partial^3 S(\rho)}{\partial\rho^3}\Big|_{\rho=\rho_0}.
    \end{eqnarray}
    In a similar fashion, we can expand the asymmetric NM incompressiblity $K(\xi)$ as 
    \begin{equation}
        K(\xi)=K+K_\tau \xi^2+{\cal{O}}(\xi^4),
    \end{equation}
    where $K$ is the incompressibility at the saturation density and
    \begin{equation}
        K_\tau= K_{sym.}-6L-\frac{Q_0 L}{K},
        \label{ktau}
    \end{equation}
      and $Q_0=27\rho^3\frac{\partial^3 \cal{E}}{\partial {\rho}^3}$ in symmetric NM at saturation density. The NM properties are compared in Table \ref{table2} for various forces used in the present calculations.
\begin{table}
\centering
\caption{The NM properties such as binding energy per particle $B/A$, $K$, effective mass ratio $M^*/M$, symmetry energy $J$ and its different co-efficients etc. are listed at the saturation density for five different parameter sets. All the parameters have MeV unit except $\rho_0$ (fm$^{-3}$) and $M^{\star}/M$ (dimensionless). The empirical / experimental values are given in the last column with their Refs. [a] \cite{Bethe_1971}, [b] \cite{Danielewicz_2014}, [c] \cite{Zimmerman_2020}, [d] \cite{Colo_2014} and [e] \cite{Stone_2014,Pearson_2010,TLi_2010}}
\renewcommand{\arraystretch}{2}
\scalebox{0.85}{
\begin{tabular}{ccccccccccc}
\hline
\hline
\multicolumn{1}{c}{Parameter}
&\multicolumn{1}{c}{NL3}
&\multicolumn{1}{c}{FSUGarnet}
&\multicolumn{1}{c}{G3}
&\multicolumn{1}{c}{IOPB-I}
&\multicolumn{1}{c}{BigApple}
&\multicolumn{1}{c}{Emp./expt.}\\
\hline
	$\rho_{0}$ (fm$^{-3})$ &  0.148  &  0.153&  0.148&0.149& 0.155&0.148--0.185 [a] \\
$B/A$  &  -16.29  & -16.23 &  -16.02&-16.10&-16.34&-15.00-- -17.00 [a]  \\
$M^{*}/M$  &  0.595 & 0.578 &  0.699&0.593&0.608&---  \\
$J$  & 37.43  &  30.95&  31.84&33.30&31.32& 30.20--33.70 [b] \\
$L$ &  118.65  &  51.04 &  49.31&63.58 &39.80&35.00--70.00 [b]\\
$K_{sym}$  &  101.34  & 59.36 & -106.07&-37.09&90.44&-174-- -31 [c] \\
$Q_{sym}$  &  177.90  & 130.93&  915.47 &862.70&1114. 74&--- \\
$K$ & 271.38  &  229.5&  243.96& 222.65 &227.00&220--260 [d]\\
$Q_{0}$ &  211.94  & 15.76&   -466.61 &-101.37&-195.67& ---\\
$K_{\tau}$ &  -703.23  &  -250.41&-307.65 &-389.46& -116.34&-840-- -350 [e] \\
\hline
\hline
\end{tabular}
\label{table2}}
\end{table}
\section{Results and Discussions}
\label{r&d}
In this section, we discuss the properties of finite nuclei, NM, and the NS matter. Some of the finite nuclei properties like binding energy per particle, charge radius, neutron-skin thickness, single-particle energy, and two neutron separation energy for few spherical nuclei are analysed. The NM parameters such as incompressibility, symmetry energy, and its different coefficients are studied for symmetric NM (SNM) and pure neutron matter (PNM). Finally, we extend our calculations to the NS and find its EOS, mass, radius, tidal deformability, and moment of inertia in detail.
\subsection{Finite Nuclei}
\subsubsection{Binding energies, charge radii, and neutron-skin thickness:-}
Here, we calculate $B/A$, charge radii $R_c$ and neutron skin thickness $\Delta r_{np}$ for eight spherical nuclei and compared with the experimental results as given in Table \ref{table3}. The BigApple parameter set well satisfies the $B/A$ and $R_c$ of the listed nuclei as compared to other parameter sets. 

\begin{figure}
\centering
\includegraphics[width=0.6\textwidth]{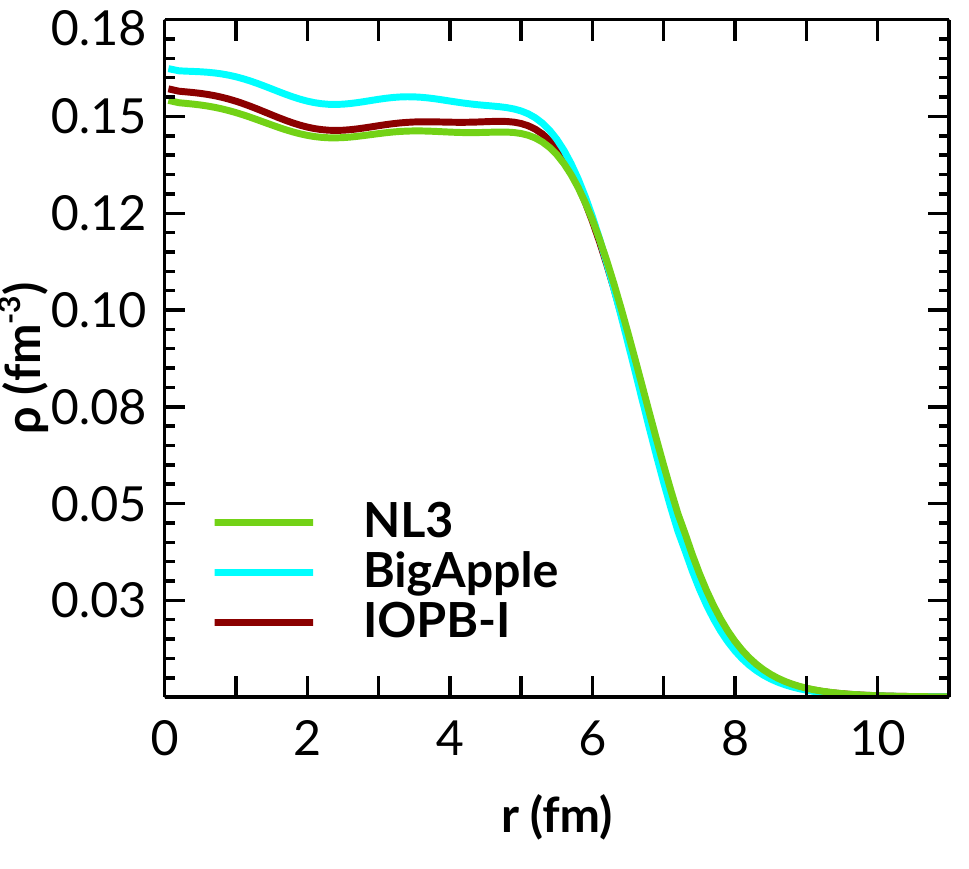}
\caption{(color online) The density of the $^{208}$Pb is shown for BigApple, NL3 and IOPB-I parameter sets.}
\label{den}
\end{figure}

We calculate the density profile of the $^{208}$Pb nucleus which is  shown in Fig. \ref{den} for BigApple with NL3 and IOPB-I parameter sets for comparison. The  central density of the nucleus larger for the BigApple case is compare to NL3 and IOPB-I. That means the nucleus will saturate at higher density for BigApple as compare to NL3 and IOPB-I. 
\begin{table} 
\centering
\caption{The numerical values of B/A (MeV) and $R_c$ (fm) and $\Delta r_{np}$ (fm) are listed with the available experimental data \cite{Wang_2012,Angeli_2013}.}
\scalebox{0.9}{
\begin{tabular}{cccccccccc}
\hline
\hline
\multicolumn{1}{c}{Nucleus}&
\multicolumn{1}{c}{Obs.}&
\multicolumn{1}{c}{Expt.}&
\multicolumn{1}{c}{NL3}&
\multicolumn{1}{c}{FSUGarnet}&
\multicolumn{1}{c}{G3}&
\multicolumn{1}{c}{IOPB-I}&
\multicolumn{1}{c}{BigApple}\\
\hline
         &B/A & 7.976 &7.917&7.876 & 8.037&7.977& 7.882 \\
$^{16}$O & R$_{c}$& 2.699 & 2.714&2.690   & 2.707&2.705&2.713  \\
         & $\Delta r_{np}$ &  & -0.026&-0.028  & -0.028&-0.027&-0.027  \\\hline
         & B/A  & 8.551 & 8.540&8.528  &8.561&8.577 &8.563 \\
$^{40}$Ca& R$_{c}$  & 3.478 &  3.466&3.438  & 3.459&3.458& 3.447 \\
         & $\Delta r_{np}$  &  & -0.046&-0.051  & -0.049&-0.049& -0.049 \\\hline
         & B/A  & 8.666 & 8.636&8.609  &8.671& 8.638& 8.547\\
$^{48}$Ca& R$_{c}$  & 3.477 & 3.443&3.426   & 3.466 &3.446& 3.447\\
         & $\Delta r_{np}$  &  &  0.229&0.169 & 0.174&0.202& 0.170 \\\hline
         & B/A  & 8.682 & 8.698&8.692  & 8.690&8.707& 8.669 \\
$^{68}$Ni& R$_{c}$  &  & 3.870 &3.861  &3.892&3.873&3.877  \\
         & $\Delta r_{np}$  &  &0.262 &0.184 &0.190&0.223&0.171  \\\hline
         & B/A  & 8.709 & 8.695&8.693  & 8.699&8.691& 8.691 \\
$^{90}$Zr& R$_{c}$  & 4.269 & 4.253&4.231  & 4.276& 4.253& 4.239\\
         & $\Delta r_{np}$  &  & 0.115&0.065   & 0.068&0.091& 0.069\\\hline
         & B/A  & 8.258 & 8.301& 8.298   & 8.266&8.284&8.259 \\
$^{100}$Sn& R$_{c}$  &  & 4.469&4.426   &4.497&4.464&4.445  \\
         & $\Delta r_{np}$  &  & -0.073&-0.078  &  -0.079&-0.077& 0.076\\\hline
         & B/A  & 8.355 & 8.371&8.372  & 8.359&8.352&8.320 \\
$^{132}$Sn& R$_{c}$  & 4.709 & 4.697&4.687   & 4.732& 4.706&4.695 \\
         & $\Delta r_{np}$  &  & 0.349&0.224   & 0.243&0.287& 0.213\\\hline
         & B/A  & 7.867 &7.885& 7.902 & 7.863&7.870& 7.894 \\
$^{208}$Pb& R$_{c}$  &5.501  & 5.509&5.496 & 5.541 &5.52& 5.495\\
         & $\Delta r_{np}$  &  &   0.283& 0.162 & 0.180 &0.221&0.151\\
\hline
\hline
\end{tabular}
\label{table3}}
\end{table}

\begin{figure}
\centering
\includegraphics[width=0.6\textwidth]{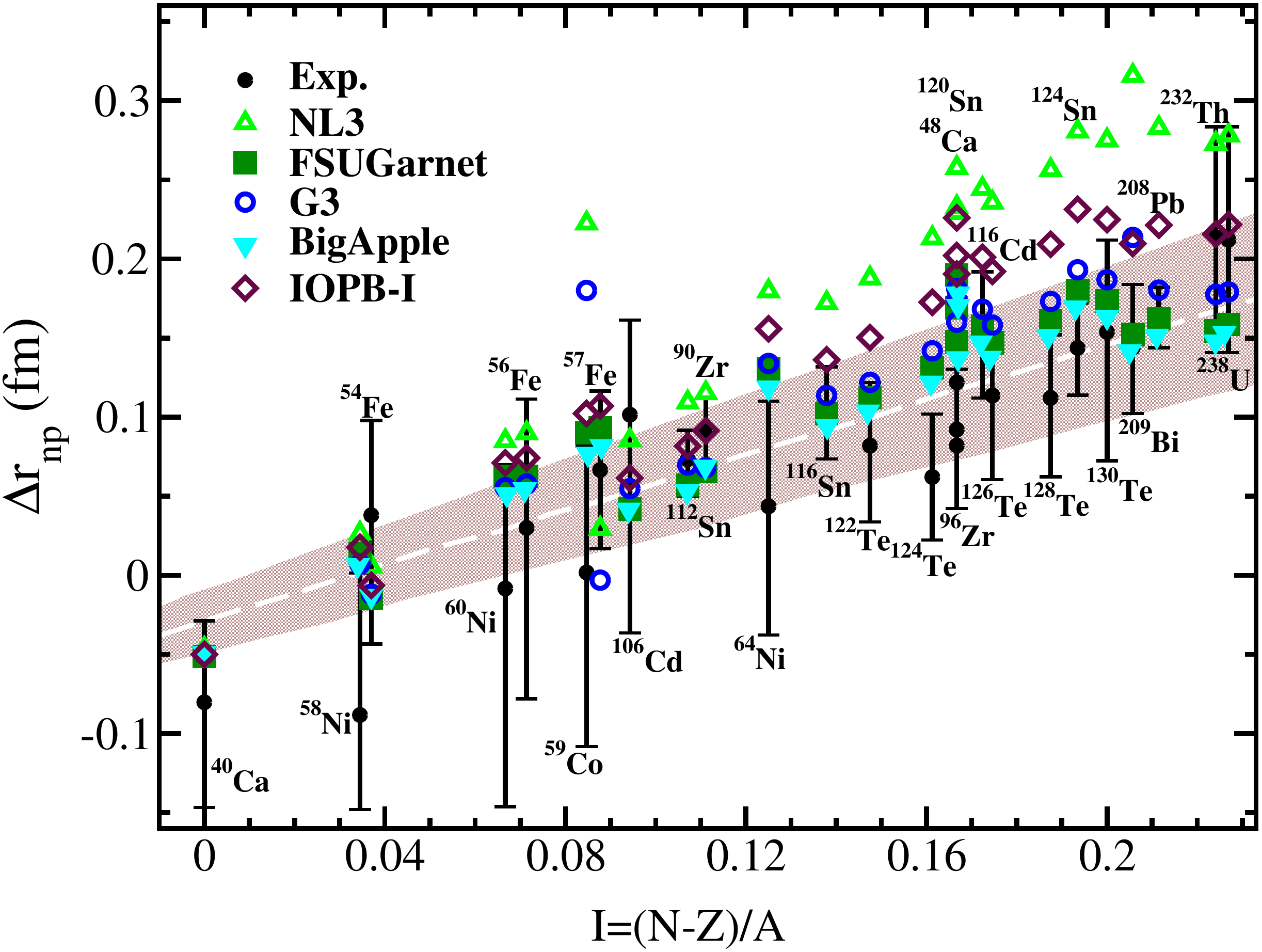}
\caption{(color online) The neutron-skin thickness as a function of the $I$. Results obtained with five different parameter sets are compared with experimental data \cite{Trzci_2001}. The shaded region is depicted using the fitting formula in Eq. (\ref{eq:rnp}).}
\label{skin}
\end{figure}

The neutron skin thickness $\Delta r_{np}$ is defined as the root mean square radii difference of neutron and proton distribution, i.e., $\Delta r_{np}=R_n-R_p$. Electron scattering experiments can determine the charge distribution of protons in the nucleus. But, it is not straightforward to calculate the neutron distributions in nuclei in a model-independent way. The Lead Radius Experiment (PREX) at JLAB has been designed to measure the neutron distribution radius in $^{208}$Pb from parity violation by the weak interaction. This measurement determined large uncertainties in the measurement of the neutron radius of $^{208}$Pb \cite{Abrahamyan_2012}. 
Recently, the PREX-II experiment gave the neutron skin thickness is \cite{Adhikari_2021}
\begin{equation}
    \Delta r_{np}=R_n-R_p=0.283\pm0.071 \ \mathrm{fm},
\end{equation}
with $1\sigma$ uncertainty. Using PREX-II data, some people have tried to constraints some NM and NS properties, which improve the understanding of the EOS for NM and NS \cite{Reed_2021, Pattnaik_2021}. Only NL3 and IOPB-I parameter set well reproduce the neutron skin thickness of the lead nucleus, which is consistent with PREX-II data. On the other hand, the neutron-skin thickness of 26 stable nuclei starting from $^{40}$Ca to $^{238}$U has been deduced by using anti-protons experiment from the low Energy anti-proton ring at CERN \cite{Trzci_2001}. The numerically calculated results and experimental data with an error bar are shown in Fig. \ref{skin}. 
\begin{equation}
    \Delta r_{np} = (0.90\pm 0.15)I+(-0.03\pm 0.02) \ {\rm fm}.
    \label{eq:rnp}
\end{equation}
The fitted data for $\Delta r_{np}$ using Eq. (\ref{eq:rnp}) are put as a band in Fig. \ref{skin}. The calculated skin-thickness of the 26 nuclei for the BigApple parameter set matches well with other sets. The skin-thickness for $^{208}$Pb nuclei with BigApple is 0.151 fm, which lies in the range given by the proton elastic scattering experiment \cite{Zenihiro_2010},  $\Delta r_{np}$ = 0.148--0.265 fm.

From Fig. \ref{skin}, we observe that the $\Delta r_{np}$ by different parameter sets coincide with each other and also with the experimental data for the nuclei with zero isospins, like $^{40}$Ca. But as the isospin asymmetry increases, the results from different parameter sets diverge from each other. Some stiff EOS like NL3, which gives a large maximum allowed mass for the NS, shows a serious divergence from the experimental data and lies outside the fitted region. 
\subsubsection{Single-particle energy:-}
\label{single}
\begin{figure}
    \centering
    \includegraphics[width=0.5\textwidth]{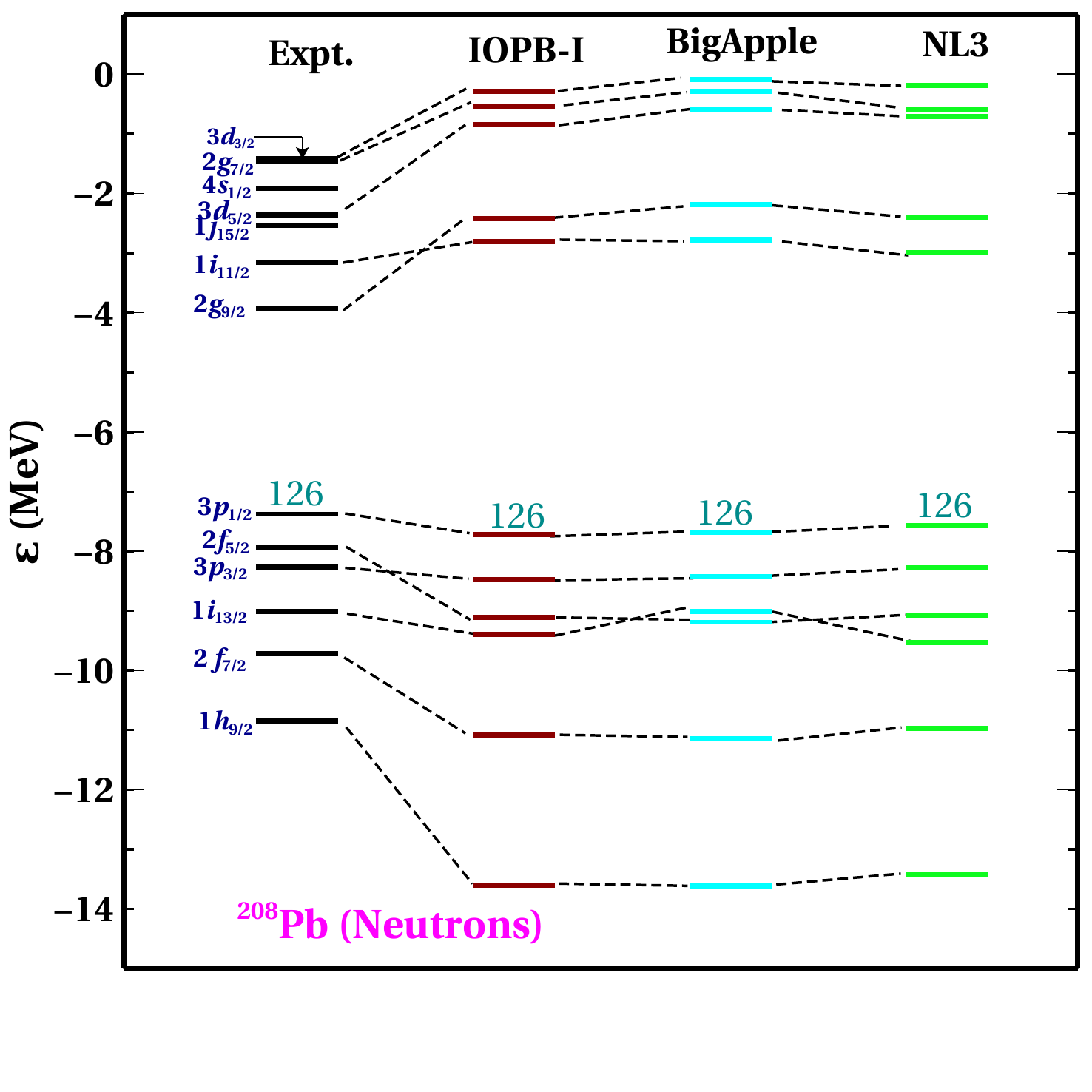}%
    \includegraphics[width=0.5\textwidth]{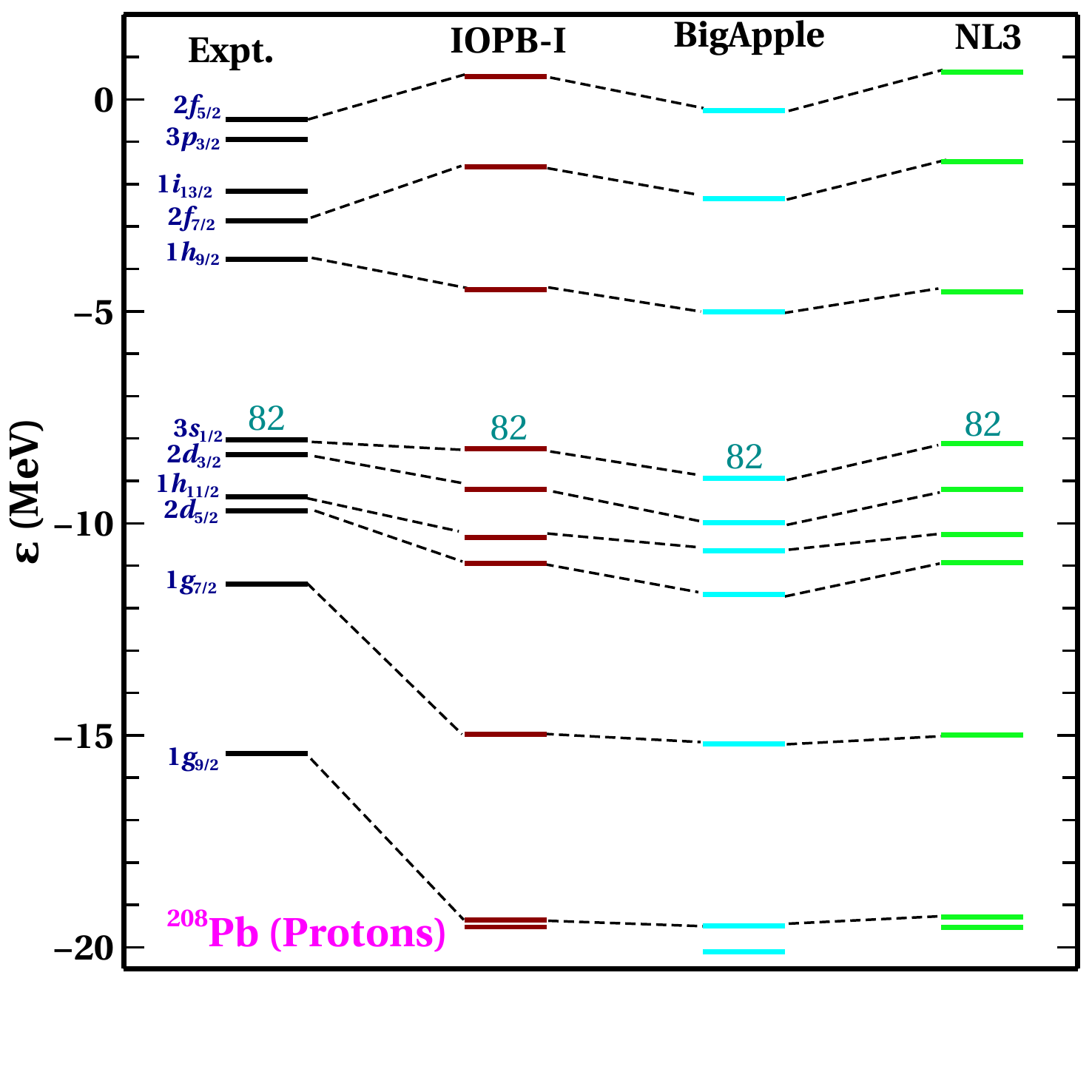}
    \caption{(colour online) The single-particle energies of $^{208}$Pb for IOPB-I, BigApple and NL3 are compared with experimental data \cite{Vautherin_1972}. The last occupied level is also shown with the numbers 126 for neutrons and 82 for protons.}
    \label{pb}
\end{figure}

\begin{figure}
    \centering
    \includegraphics[width=0.5\textwidth]{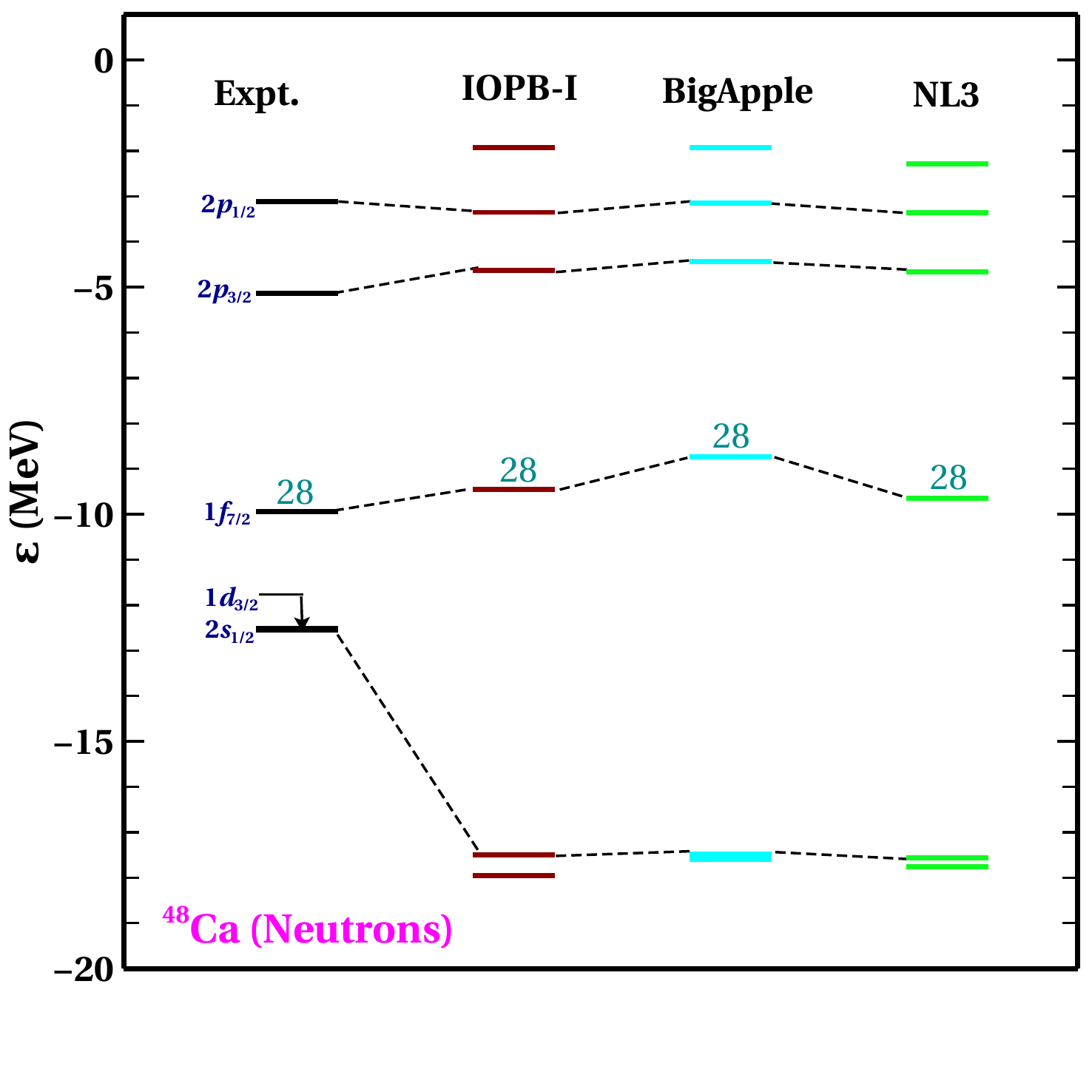}%
    \includegraphics[width=0.5\textwidth]{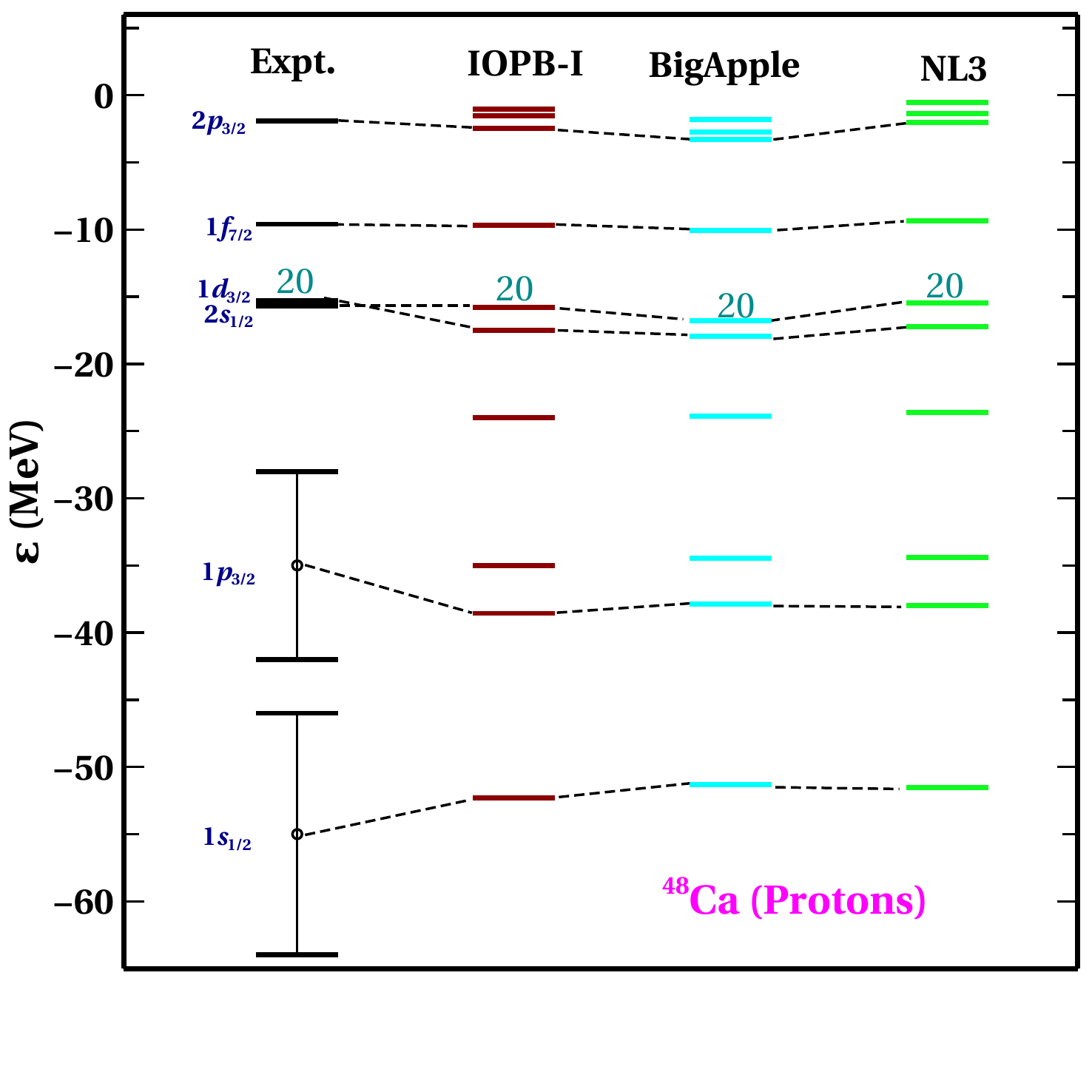}
    \caption{(colour online) Same as Fig. \ref{pb}, but for $^{48}$Ca.}
    \label{ca}
\end{figure}

The study of single-particle energies for nuclei gives us an indication of shell closer. From this, we can identify the large shell gaps and predict the presence of the magic numbers. Here, we calculate the single-particle energies of two doubly magic nuclei as representative cases, e.g., $^{48}$Ca and $^{208}$Pb for IOPB-I, BigApple, and NL3 parameter sets. The predicted single-particle energies for both protons and neutrons of $^{48}$Ca and $^{208}$Pb are compared with the experimental data \cite{Vautherin_1972} in Figs. \ref{pb} and \ref{ca}. The BigApple set well predicted the magicity compared to the other parameter sets. All three parameter sets reproduce the known magic numbers 20, 28, 82, and 126. The nuclei, $^{48}$Ca and $^{208}$Pb, are doubly closed, which are considered to be perfectly spherical.
\subsubsection{Two-neutron separation energy $S_{2n}(Z,N)$:-}
\begin{figure} 
\centering
\includegraphics[width=0.6\textwidth]{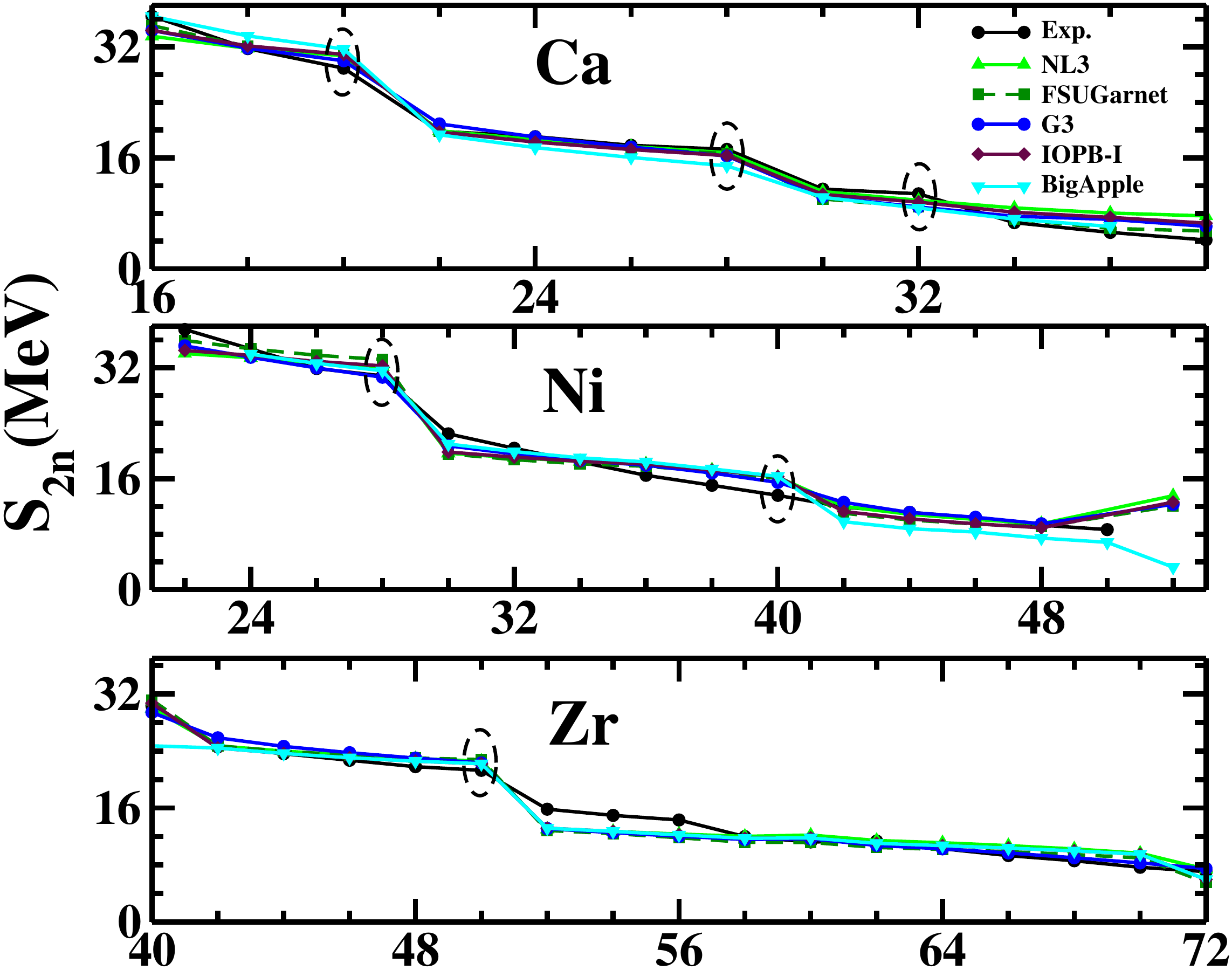}
\includegraphics[width=0.6\textwidth]{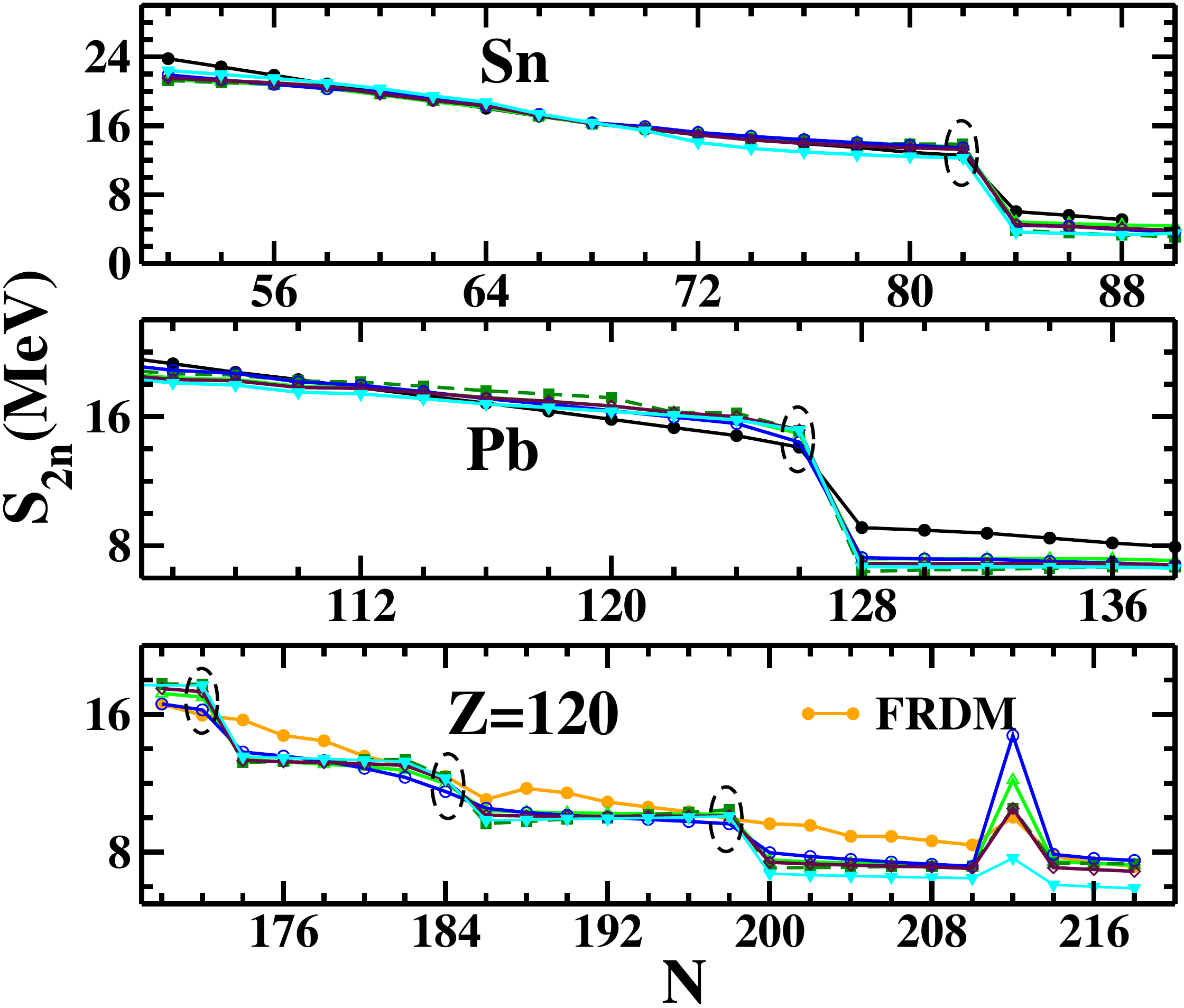}
\caption{(colour online) The two-neutron separation energy as a function of neutron number for the isotopic nuclei like Ca, Ni, Zr, Sn and Pb for five different parameter sets. The FRDM \cite{Moller_2016} data and experimental data \cite{Wang_2012} are also given for comparison. The circle represents the magicity of the nuclei.}
\label{s2n}
\end{figure}

The two neutron separation energy $S_{2n}(N,Z)$ is the energy required to remove two neutrons from a nucleus with N neutrons and Z protons, i.e. 
\begin{equation}
    S_{2n}(N,Z)=B(N,Z)-B(N-2,Z).
\end{equation}
The study of neutron separation energy is essential to explore the nuclear structure near the drip line. A sudden drop in $S_{2n}(N, Z)$ represents the beginning of a new shell. The large shell gap in single-particle energy levels indicates the magic number, and it is responsible for the extra stability for the magic nuclei. We calculate the $S_{2n}$ for six isotopic chains Ca, Ni, Zr, Sn, Pb, and $Z=120$, which are shown in Fig. \ref{s2n} and compared with experimental data given in the Ref. \cite{Wang_2012}. We also compare the results obtained for the $Z=120$ isotopic chain results with the finite range droplet model (FRDM) \cite{Moller_2016}. From Fig. \ref{s2n}, it is cleared that the value of $S_{2n}$ decreases with the increase of the neutron number, i.e., towards the neutron drip line. All the magic characters appear at neutron number $N=20, 28, 32, 40, 50, 82, 126$. In the last part of Fig. \ref{s2n}, the magicity are found at $N=172, 184, 198$ for $Z=120$ nuclei. In this case, we compare the calculated data with FRDM \cite{Moller_2016}, there is no experimental data available for the Z=120. There is a sharp fall in the $S_{2n}$ for five different parameter sets, which are consistent with the prediction of various models in the superheavy mass region \cite{Rutz_1997, Gupta_1997, Patra_1999, Mehta_2015}. Bhuyan {\it et al.} \cite{Bhuyan_2012} have predicted that $Z=120$ is next the magic number after $Z=82$, which lies in the superheavy region. Also, Mehta {\it et al.} \cite{Mehta_2015} have predicted that $Z=120$ nuclei are spherical in their ground state, and possible proton magic number at $Z=120$. We hope the future experiments may answer the shell closure at $N=172, 184$ and $198$.
\subsubsection{Isotopic shift}
The Isotopic shift is defined as, $\Delta r_c^2= R_c^2(208)-R_c^2(A)$, where we take $^{208}$Pb as reference nucleus. In Fig \ref{s2n}, we plot the $\Delta r_c^2$ for Pb isotopes for five parameter sets. The experimental data are also given for comparison. The predicted $\Delta r_c^2$ by BigApple set well match with NL3 set. The iso-spin-dependent term in the nuclear interaction results in the kink in the isotopic shift graph.  In the conventional RMF model, the spin-orbital term is included automatically by assuming nucleons as the Dirac spinor. But the situation is not the same in the Skyrme model where one has to add the spin-orbital part to match with the experimental results \cite{Sharma_1995}.
\begin{figure}
\centering
\includegraphics[width=0.8\textwidth]{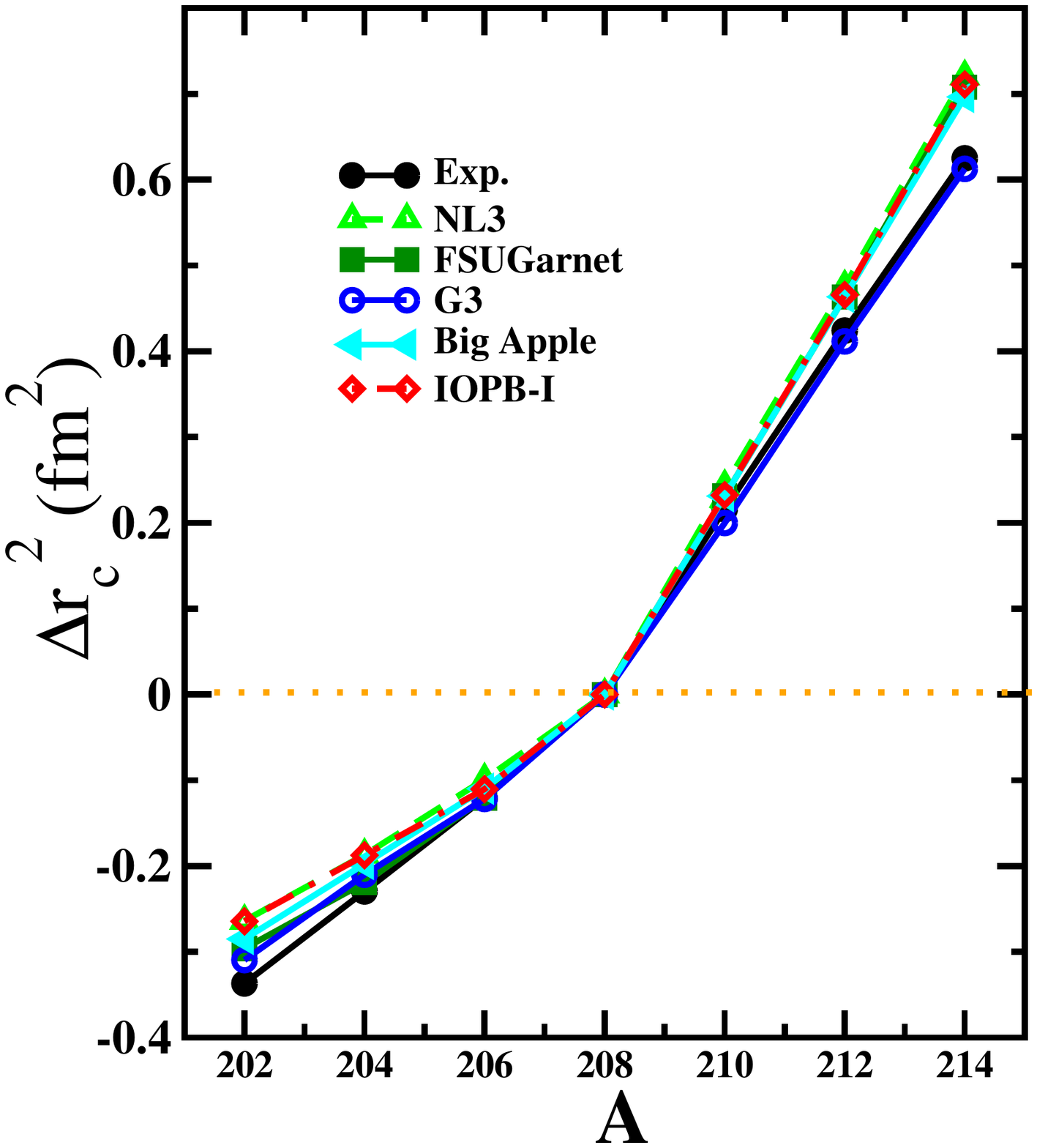}
\caption{Isotopic shift for Pb isotopes for five parameter sets is shown taking $^{208}$Pb as reference.}
\label{isotopic}
\end{figure}

In conclusion, we calculate some finite nuclei properties such as $B/A$, $R_c$, $\Delta r_{np}$, single-particle energy, $S_{2n}$, $\Delta r_c^2$ for BigApple parameter set along with other four sets. We find that the $B/A$, $R_c$ for some nuclei are well reproduced by the BigApple like other parameter sets. The skin thickness of lead nuclei is found to be 0.151 fm for BigApple set, which is inconsistent with the PREX-II data \cite{Adhikari_2021} but it satisfies the CERN data \cite{Trzci_2001}. The single-particle energy for $^{48}$Ca and $^{128}$Pb are well reproduced by the BigApple set. The BigApple set well predicts the two neutron separation energies for series nuclei, including $Z=120$ as compared to other sets. Finally, the isotopic shift is almost well consistent with experimental data. From the above studies, we conclude that one can take BigApple set to calculate finite nuclei properties.
\subsection{Nuclear Matter}
In this sub-section, we study NM parameters like $B/A$, incompressibility $K$, density-dependent symmetry energy $S(\rho)$ and its different coefficients like slope $L$, curvature $K_{sym}$, skewness $Q_{sym}$, etc. in detail. Here we give a  special emphasis on the newly developed parameter set BigApple \cite{Fattoyev_2020}. The values of NM quantities are given in Table \ref{table2}. First, we discussed the incompressibility of the NM. For BigApple, the value of $K=227$ MeV, which lies in the experimental data range obtained from the excitation energy of the isoscalar giant monopole resonances of $^{208}$Pb and $^{90}$Zr  and its value is $K=240\pm20$ MeV \cite{Colo_2014,Stone_2014}. Recently the value of $K_{sym}$ is constrained by Zimmerman $\textit{et al.}$ \cite{Zimmerman_2020} combining the GW170817 and NICER data and it is found to be $102_{-72}^{+71}$ MeV at $1\sigma$ level. The values of symmetry energy and its slope for BigApple are 33.32, and 39.80 MeV, respectively, which are also lie in the range given by Danielewicz and Lee \cite{Danielewicz_2014} at the saturation density (see Table \ref{table2}). 
\begin{figure}
\centering
\includegraphics[width=0.7\textwidth]{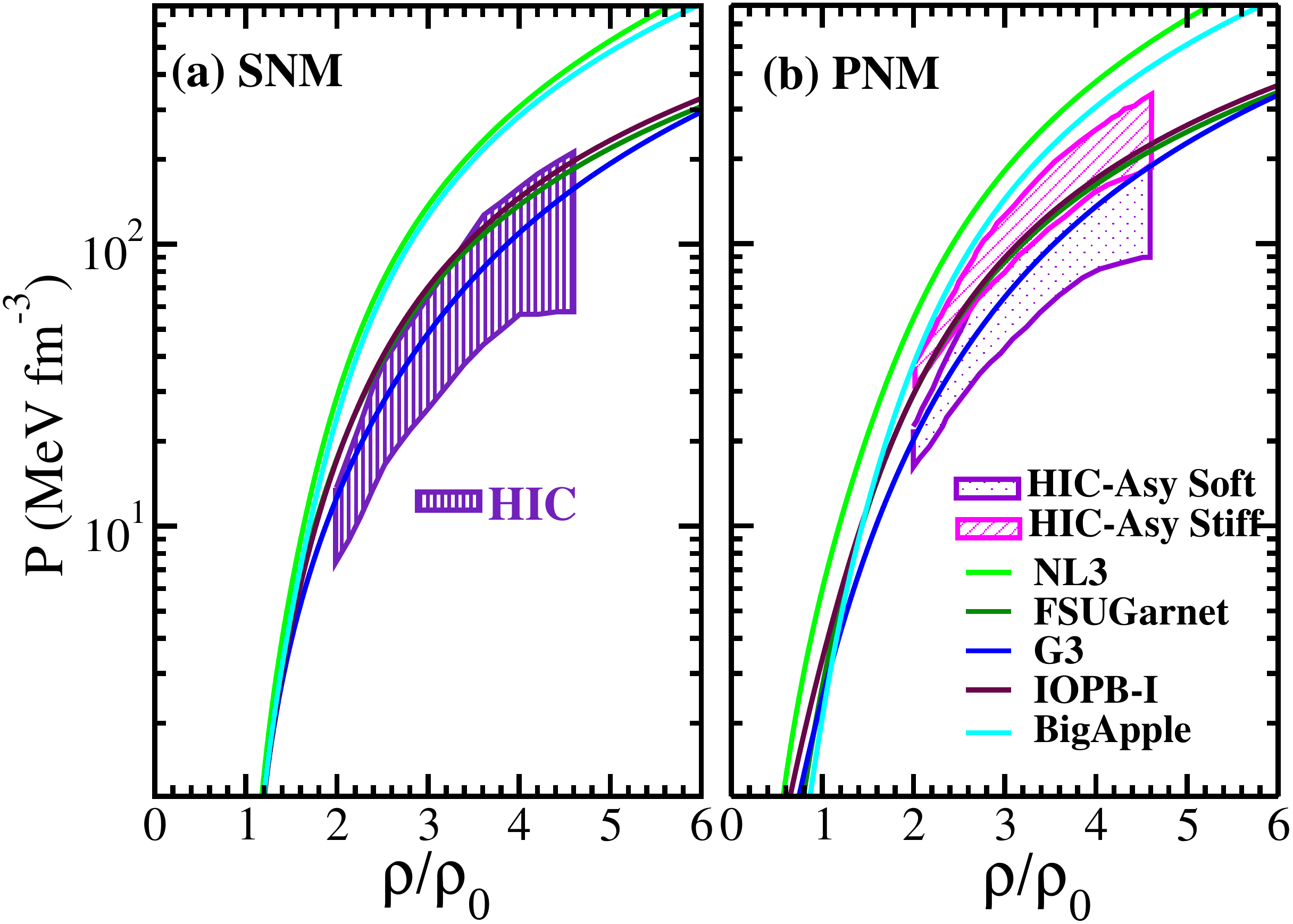}
\caption{(colour online) Calculated pressure with the variation of baryon density. The results for NL3, BigApple, FSUGarnet, G3 and IOPB-I are compared with HIC data \cite{Danielewicz_2002} both for symmetric nuclear matter (left) and pure neutron matter (right). For the PNM case, the same data is divided into (i) HIC-Asy soft (ii) HIC-Asy stiff, mainly based on the density dependence symmetry energy.}
\label{EOSnm}
\end{figure}

In Fig. \ref{EOSnm}, we plot the pressure with the variation of the baryon density for SNM and PNM and compared with the experimental flow data \cite{Danielewicz_2002}. The calculated pressure by the G3 set is consistent with heavy-ion collisions (HIC) data for the whole densities range for SNM (shown in Fig. \ref{EOSnm}). Although the parameter sets like IOPB-I and FSUGarnet reproduce stiffer EOS compared to G3, their calculated pressure still matches HIC data. NL3 and BigApple are the stiffer EOSs as compared to others. So they disagree with the HIC data both for SNM and PNM cases. Although the EOS corresponds to the BigApple parameter set doesn't pass through the experimental shaded regions given by HIC; still, it predicts the value of $K$, $J$, and $L$, which is reasonably within the empirical/experimental limit as given in Table \ref{table2}.

\begin{figure}
\centering
\includegraphics[width=0.6\textwidth]{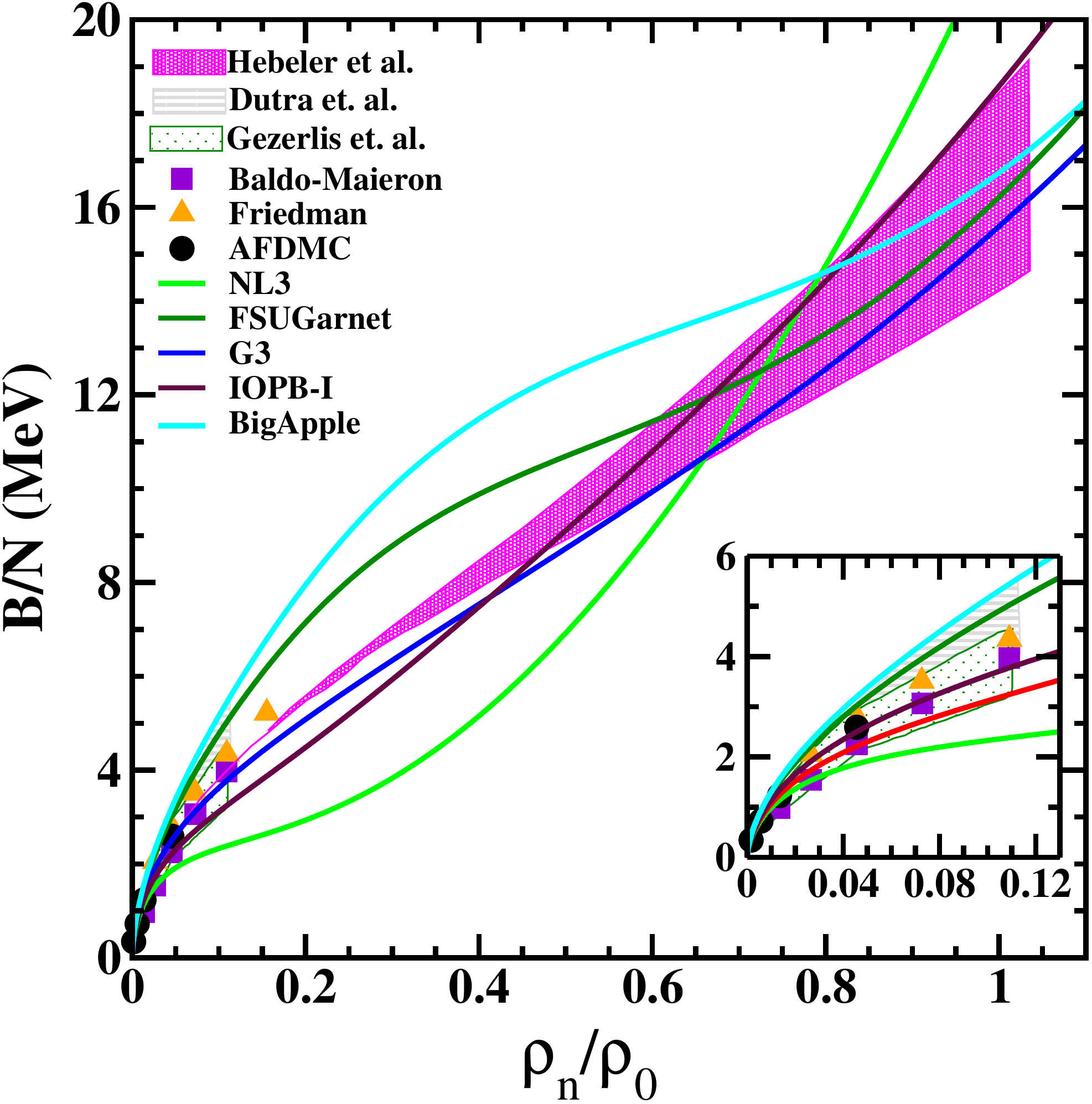}
\caption{(color online) The $B/N$ as a function of density with  NL3 \cite{Lalazissis_1997}, FSUGarnet \cite{Chen_2015}, G3 \cite{Kumar_2017}, IOPB-I \cite{Kumar_2018} and BigApple \cite{Fattoyev_2020} parameter sets. The other results are from Hebeler {\it et al.} \cite{Hebeler_2013}, Dutra {\it et al.} \cite{Dutra_2012}, Gezerlis {\it et al.} \cite{Gezerlis_2010}, Baldo-Maieron \cite{Baldo_2008}, Friedman \cite{Friedman_1981} and Auxiliary-field diffusion Monte Carlo (AFDMC) \cite{Gandolfi_2008}.}
\label{be}
\end{figure}

Next, our focus is on the $B/A$ at the saturation density. The variation of $B/A$ with baryon density ($\rho/\rho_0$) for the PNM system is shown in Fig. \ref{be} for BigApple parameter set along with NL3, FSUGarnet, G3, and IOPB-I. Some experimental data are also put for comparison. From this plot, one can see that at the low-density regions, except BigApple and NL3, other parameter sets are in harmony with the results of the microscopic calculations. The BigApple, G3, IOPB-I, and FSUGarnet parameter sets pass through the shaded regions near the saturation density. It means that these parameter sets are qualitatively consistent with the results obtained by Hebeler {\it et al.} data \cite{Hebeler_2013}. The $B/A$ at the saturation density for considered parameter sets lies in the empirical limit as given in Table \ref{table2}.
\begin{figure}
\centering
\includegraphics[width=0.6\textwidth]{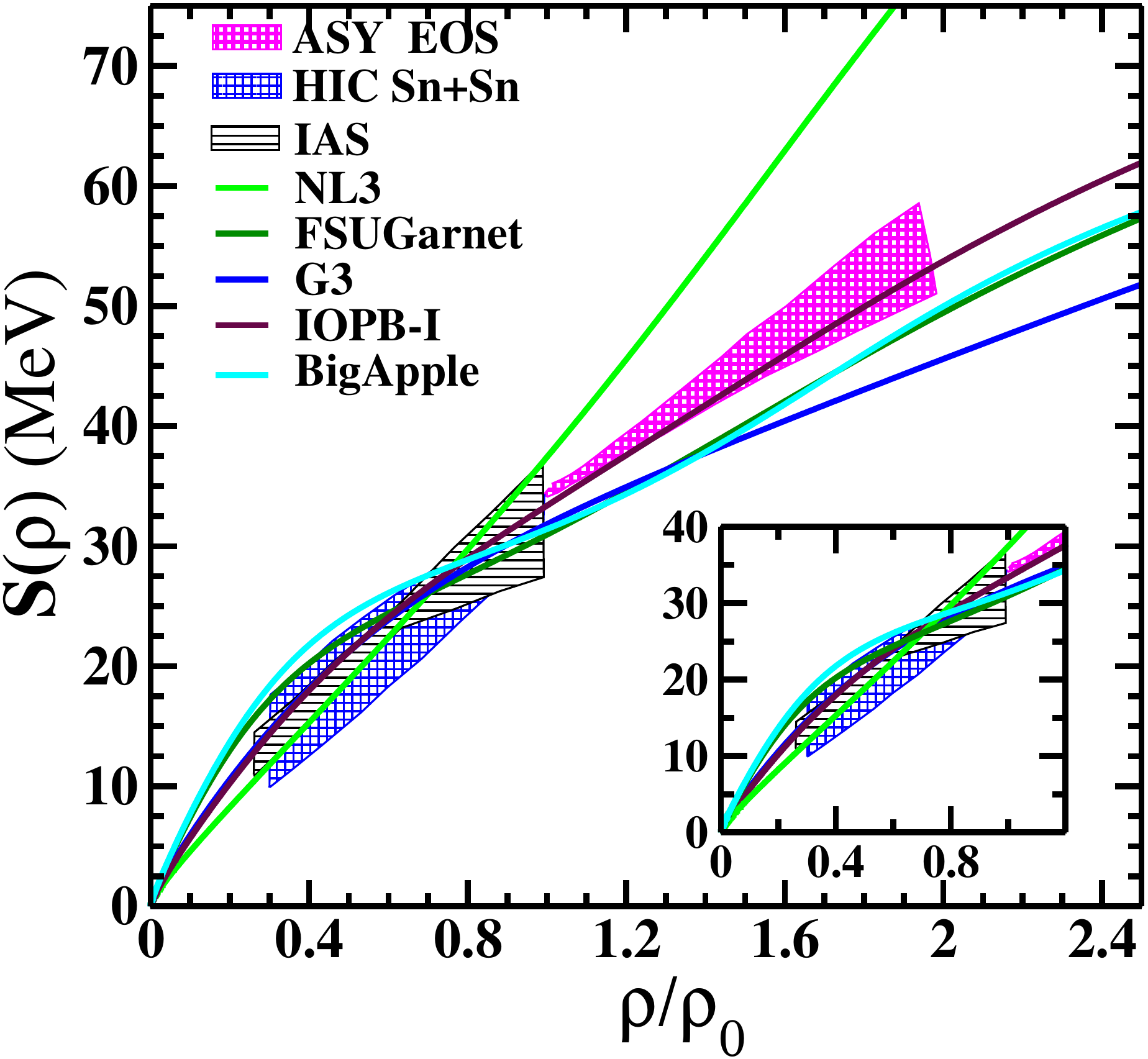}
\caption{(color online) The density dependent symmetry energy with the baryon density for five different parameter sets . The shaded region is the symmetry energy from IAS \cite{Danielewicz_2014}, HIC Sn+Sn \cite{Tsang_2009,Tsang_2010} and ASY-EOS experimental data \cite{Russotto_2016}. The zoomed pattern of the symmetry energy at low densities is shown in the inset.}
\label{sym}
\end{figure}

The symmetry energy is defined as the difference between energy per particle of PNM and SNM. The value of the symmetry energy at saturation density is known up to some extent, but its density dependence variation is still uncertain, i.e., the value of the symmetry energy at the saturation density is better constrained than its density dependence. It has diverse behavior at the different densities regions \cite{BaoLi_2019}. Also, the symmetry energy has broad relations with some properties of the NS \cite{Dorso_2019,Zhang_2020,Lattimer_2014,Gandolfi_2016}.

In Fig. \ref{sym}, we show the density-dependent symmetry energy with baryon density for five parameter sets. The symmetry energy for G3, IOPB-I, and FSUGarnet predict soft behavior at the low density due to cross-coupling between $\omega$ and $\rho$ meson. But for the BigApple case, the values of $S(\rho)$ at low-density regions are too high. Also, at higher density, it predicts softer $S(\rho)$, which doesn't pass through the ASY data \cite{Russotto_2016}. This shows a poor density dependence of the symmetry energy for the BigApple case.

In summary, we check the status of the stiff EOSs like NL3 and BigApple parameter sets to reproduce different constraints given by pure nuclear matter (PNM), B/A, and $S(\rho)$,  which are given shown in Figs. \ref{EOSnm}--\ref{sym} respectively. We find that the BigApple set doesn't satisfy those constraints such as flow data, symmetry energy constraints due to (i) Its stiff behavior at low-density regions. (ii) The saturation density (0.155 fm$^{-3}$) for the BigApple parameter set is more as compared to the other two parameter sets, see Fig. \ref{den}. The density distributions for NL3 and IOPB-I are almost the same, but there is a substantial shift for the BigApple case at the center. Hence in this sub-section, we examine the predictive capacity of the BigApple parameter set, i.e., we check how safe to take the BigApple parameter set for the study of NM properties. We find that it doesn't look too safe to take the BigApple set for the calculation.
\subsection{Neutron Star}
\subsubsection{Equation of state of the NS:-}
\begin{figure}
\centering
\includegraphics[width=0.6\textwidth]{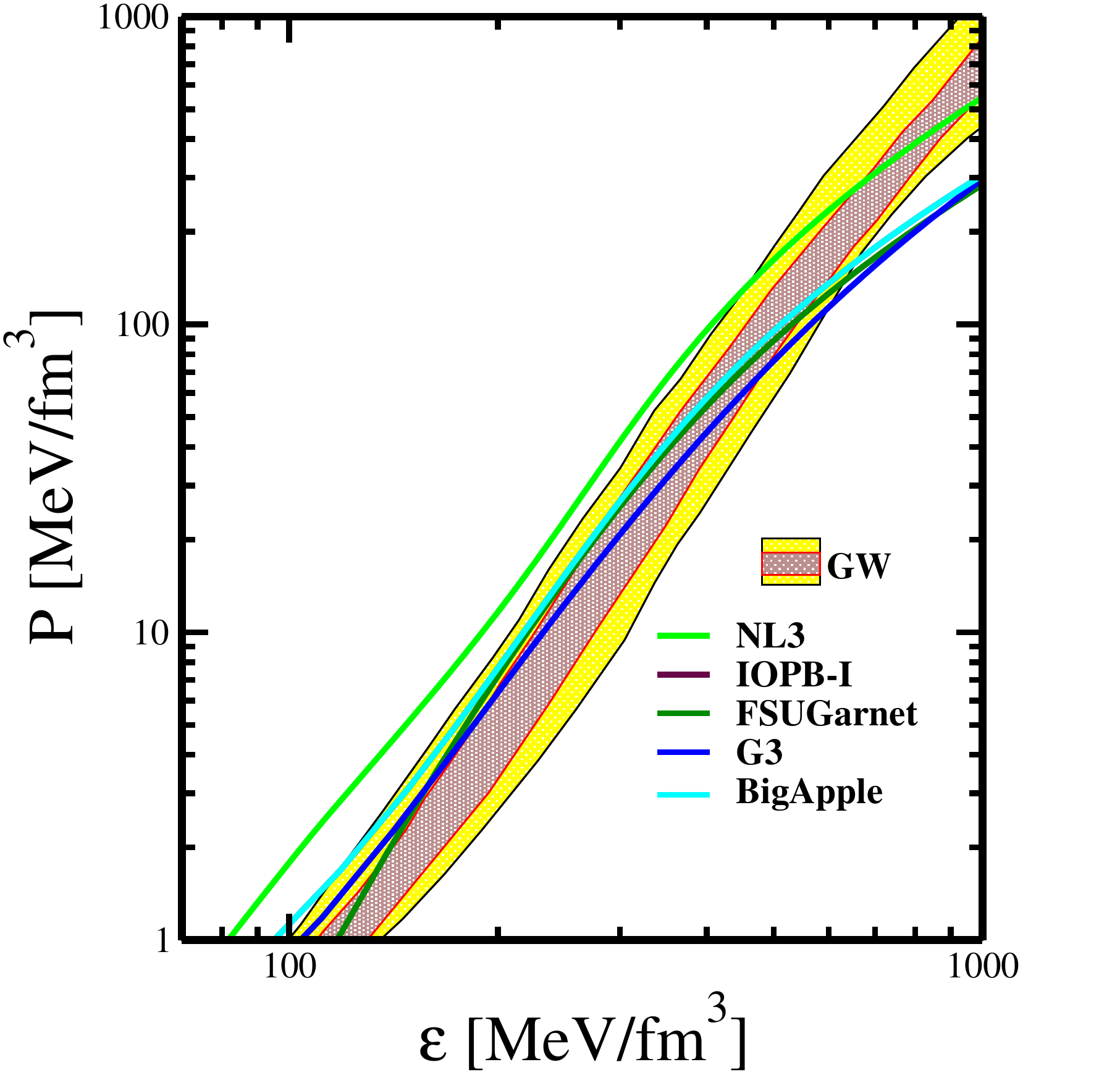}
\caption{(color online) The equations of states of $\beta$-equilibrated matter for NL3, FSUGarnet, G3, IOPB-I and BigApple parameter sets. The shaded regions are for 50 \% (orange) and 90\% (grey) posterior credible limit given by the GW170817 data \cite{Abbott_2018}.}
\label{EOSns}
\end{figure}

Nuclear EOS is the main ingredient to study NS properties. The predicted EOS for five models alongside extracted recent GW170817 observational data are shown in Fig.  \ref{EOSns}. The shaded regions are deduced from GW170817 data with 50\% (grey) and 90\% (yellow) credible limit  \cite{Abbott_2018}. For the crust part, we use the BCPM crust EOS \cite{BKS_2015} and join it with the uniform liquid core to form a unified EOS. Having the EOS, one can now proceed in the next section to solve Tolman-Oppenheimer-Volkoff (TOV) \cite{TOV1, TOV2} to calculate the properties of the NS.

\begin{figure}
\centering
\includegraphics[width=0.6\textwidth]{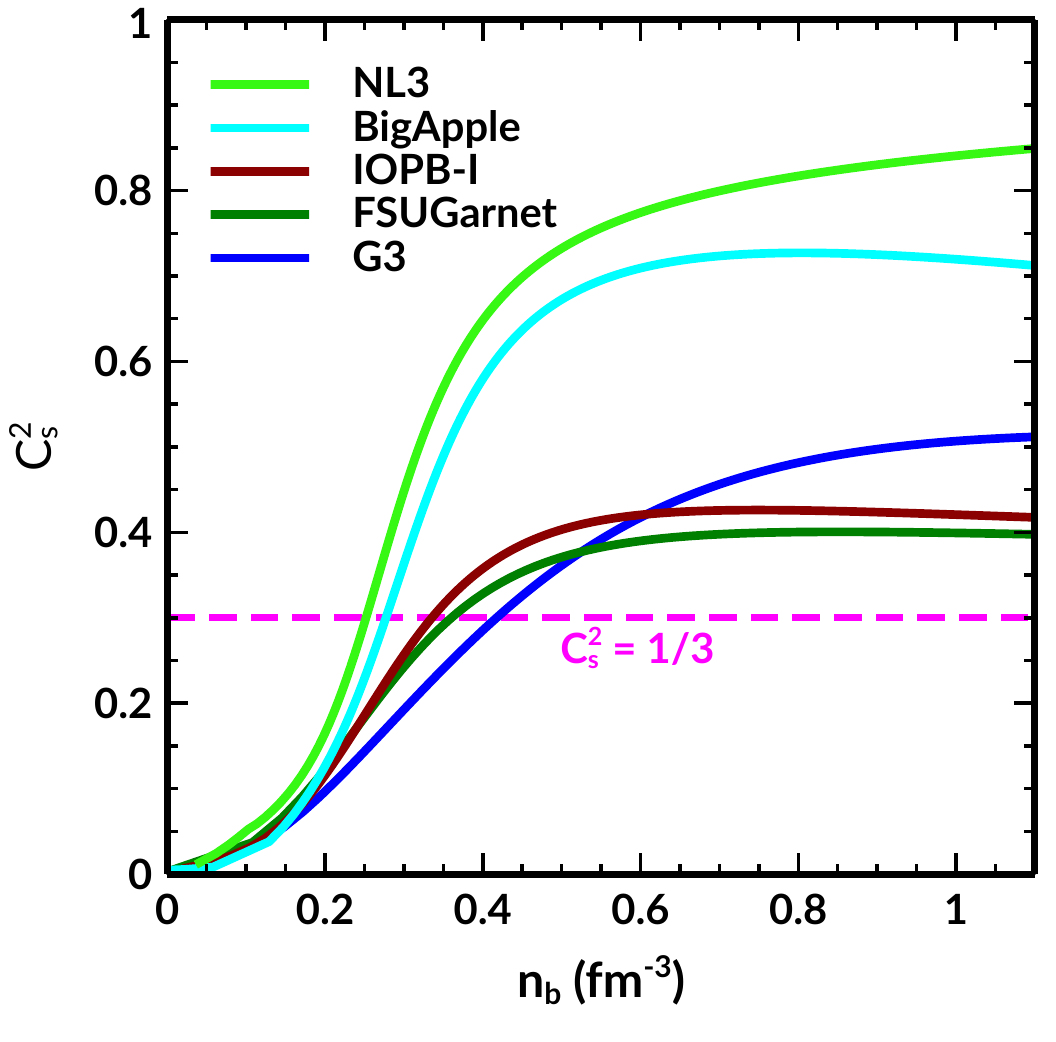}
\caption{(color online) Sound speed as a function of baryon density ($n_b$) for five considered parameter sets. The dashed magenta line represents the QCD conformal limit ($C_s^2=1/3$)}.
\label{s_speed}
\end{figure}

We calculate the sound speed inside the NS using the equation $C_s^2=\partial p/\partial e$. We plot the variation of $C_s^2$ as a function of baryon density ($n_b$) in Fig. \ref{s_speed}. All the parameter sets respect causality in the whole density region. The value of $c_s^2$ increases up to 0.4 fm$^{-3}$, and it becomes constant beyond that, which is consistent with Fattoyev {\it et al.} \cite{Fattoyev_2020}. 
\subsubsection{Mass, radius, tidal deformability and moment of inertia of the neutron star:-}
Here, we calculate the mass, radius, and tidal deformability of a non-rotating NS. A star is deformed in the field created by its companion star. The induced mass quadruple moment $Q_{ij}$ has a linear relationship with external tidal field $\mathcal{E}_{ij}$ by a proportionality constant $\lambda$ is called tidal deformability.  \cite{Hinderer_2010, Kumartidal_2017}
\begin{eqnarray}
 Q_{ij}=-\lambda\mathcal{E}_{ij},\ \mathrm{where} \  \lambda=\frac{2}{3} k_2 R^{5},
\label{eqtidal1}
\end{eqnarray}
and
\begin{eqnarray}
\Lambda=\frac{\lambda}{M^5}=\frac{2k_2}{3C^5},
\label{eqtial2}
\end{eqnarray}
 where $k_2$ and $\Lambda$ are the second Love number and dimensionless tidal deformability. $C$ is the compactness parameter ($C=M/R$). To calculate the Love number one has to solve the following differential equation with condition $y\equiv y(R)$ for quadrupole case ($l=2$) \cite{Hinderer_2008,Kumartidal_2017} 
\begin{figure}
\centering
\includegraphics[width=0.6\textwidth]{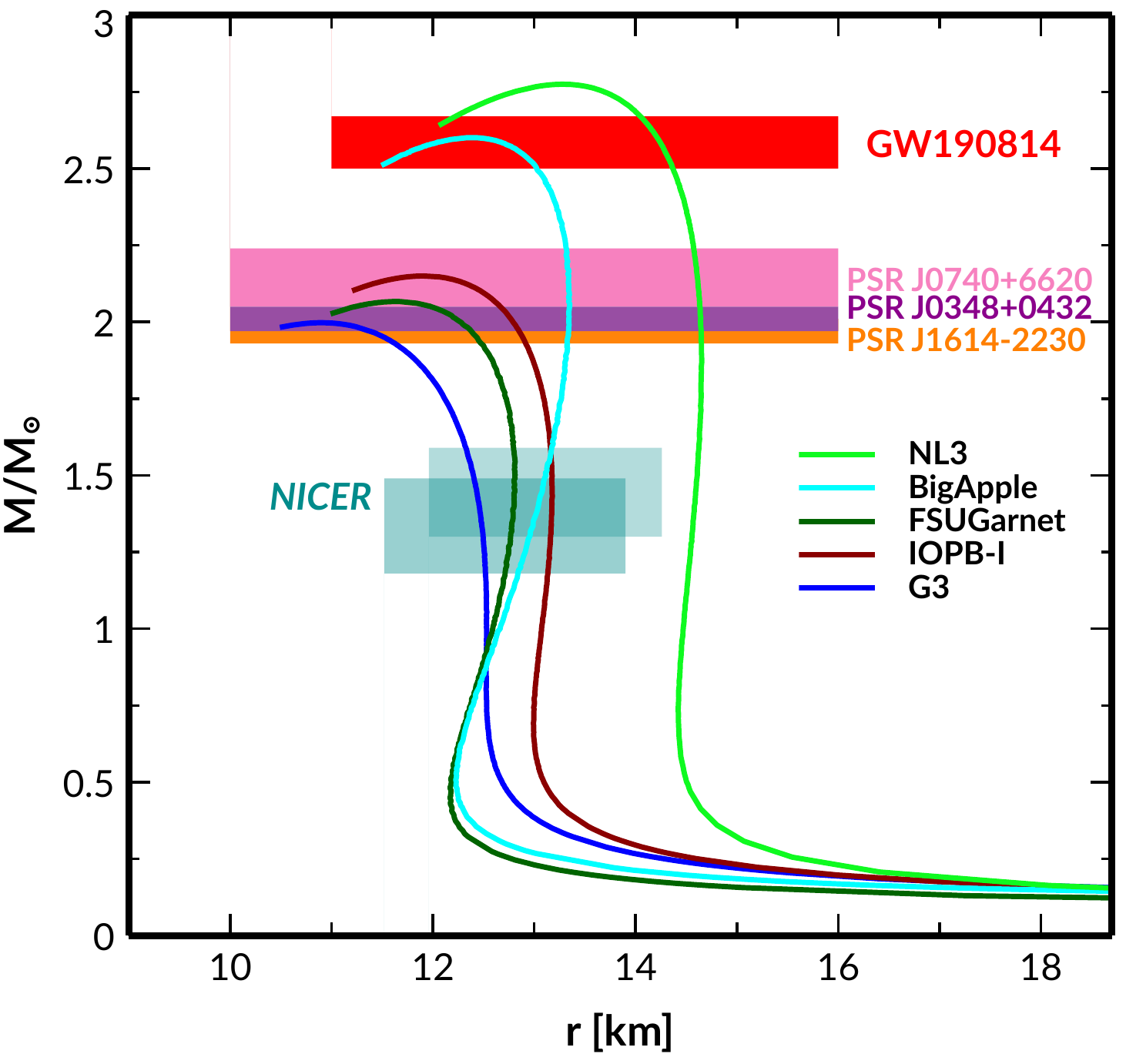}
\caption{(color online) The mass-radius profile predicted by NL3, FSUGarnet, G3, IOPB-I and BigApple. The different colour bands shows the mass of the NS observed from the different pulsars. The mass constraint by GW190814 \cite{RAbbott_2020}, and NICER results from two different analysis \cite{Miller_2019, Riley_2019} are also depicted.}
\label{MR}
\end{figure}
\begin{eqnarray}
r\frac{d y(r)}{d r} + y(r)^{2} + y(r) F(r) + r^{2} Q(r)=0,
\label{eqtidal3}
\end{eqnarray}
where
\begin{eqnarray}
 F(r)&=&\frac{r - 4 \pi r^{3} [{\mathcal{E}}(r)-P(r)]}{r-2 M(r)},\\
 Q(r)&=&\frac{4 \pi r(5{\mathcal{E}}(r)+9 P(r)+\frac{{\mathcal{E}}(r)+P(r)}
{{\partial P(r)}/ {\partial{\mathcal{E}}(r)}}-\frac{6}{4 \pi r^{2}})}{r-2 M(r)}
\nonumber \\
&&
-4 \Big[\frac {M(r)+4 \pi r^{3} P(r)}{r^{2}(1- 2 M(r)/r)}\Big]^{2}.
\end{eqnarray}
The Love number expression is given as \cite{Hinderer_2010}
\begin{eqnarray}
k_{2} &=& \frac{8}{5}(1-2C)^{2}C^{5}[2C(y-1)-y+2]
\Big\{ 2C(4(y+1)C^{4} \nonumber \\
&&
+(6y-4)C^{3}+(26-22y)C^{2}+3(5y-8)C-3y+6)\nonumber\\
&&
-3(1-2C)^{2}(2C(y-1)-y+2)log\Big( \frac{1}{1-2C}\Big)\Big\}^{-1}.\nonumber\\
\label{eqlove}
\end{eqnarray}
For a realistic star the value of $k_2$ is 0.05--0.1 \cite{Hinderer_2008}. Hence, the $\lambda$ of a star can be obtained by integrating Eq. (\ref{eqtidal3}) simultaneously with TOV equations for a given EOS \cite{TOV2}:
\begin{eqnarray}
\frac{d P(r)}{d r}=-\frac{[{\mathcal{E}}(r)+P(r)][M(r)+{4\pi r^3 P(r)}]}{r^2(1-\frac{2M(r)}{ r})}, 
\label{eqtov1}
\end{eqnarray}
and
\begin{eqnarray}
\frac{d M(r)}{d r}={4\pi r^2 {\mathcal{E}}(r)}.
\label{eqtov2}
\end{eqnarray}
Starting with the initial boundary conditions at $P(0) = P_{c}$, $M(0)=0$, and $y(0)=2$; and at the surface of the star the boundary conditions are $P(R)=0$, $M(R)=M$ and $r(R)=R$. We solve Eqs. (\ref{eqtidal1}-\ref{eqtov2}) for a given EOS of the star. Therefore, for the given central density one can uniquely determined the $M$, $R$ and $k_2$ for an isolated NS.
\begin{table*}
\centering
\caption{The maximum mass $M_{max}$, central density ${\cal{E}}_c$, radius $R$, dimensional tidal deformability $\Lambda$, compactness $C$ and dimensionless canonical moment of inertia $\bar{I}_{1.4}$ are given both for canonical (1.4) and maximum mass NS for BigApple along with other four different parameter sets.}
\label{table4}
\scalebox{0.84}{
\begin{tabular}{lllllllllllll}
\hline \hline
\multirow{2}{*}{\begin{tabular}[c]{@{}l@{}}\hspace{0.2cm}Model\end{tabular}} &
\multirow{2}{*}{\begin{tabular}[c]{@{}l@{}}$M_{max.}$\\($M_{\odot}$)\end{tabular}} &
\multicolumn{2}{l}{\begin{tabular}[c]{@{}l@{}} \hspace{0.5cm}${\cal{E}}_c$\\\hspace{0.2cm}(MeV/fm$^3$)\end{tabular}} &
\multicolumn{2}{l}{\begin{tabular}[c]{@{}l@{}}\hspace{0.5cm}$R$\\ \hspace{0.5cm}(km)\end{tabular}} &
\multicolumn{2}{l}{\begin{tabular}[c]{@{}l@{}}\hspace{0.7cm}$\Lambda$\\ \end{tabular}} &
\multicolumn{2}{l}{\begin{tabular}[c]{@{}l@{}}\hspace{0.5cm} $C$\\ \end{tabular}}& 
\multicolumn{2}{l}{\multirow{2}{*}{\hspace{0.2cm}$\bar{I}_{1.4}$}} \\ 
& & 1.4 & max. & 1.4 & max. & 1.4 & max. & 1.4 & max.& \multicolumn{2}{l}{}  \\ \hline 
NL3  &2.77&270 & 870 & 14.58 & 13.28 & 1267.79& 4.49 & 0.141 & 0.308 &\multicolumn{2}{l}{16.970}\\ \hline
BigApple &2.60 & 326&980&12.96&12.41& 717.30&5.00&  0.159   & 0.308 &\multicolumn{2}{l}{14.538}    \\ \hline
IOPB-I &2.15 & 366 & 1100& 13.17 & 11.91 & 681.27&14.82 & 0.156 &  0.265 & \multicolumn{2}{l}{14.278}    \\ \hline
FSUGarnet &2.07  & 384  & 1120     &  12.87   & 11.71     &  624.81   & 18.20     & 0.161    &  0.260 &\multicolumn{2}{l}{13.940}   \\ \hline
G3 &1.99 &460   & 1340     & 12.46    &  10.93    & 461.28    & 12.16     & 0.165    &  0.270 &\multicolumn{2}{l}{12.857}  \\ \hline \hline
\end{tabular}}
\end{table*}

The mass and radius of the NS for five different parameter sets are displayed in Fig. \ref{MR}, alongside the recent observational limit on the maximum mass of the NS. All the five EOSs are able to support maximum NS mass 2.0 $M_{\odot}$, which are compatible with the constraint of precise measured NS masses from PSR J1614-2230 \cite{Demorest_2010}, PSR J0348+0432  \cite{Antoniadis_2013}, PSR J0740+6620 \cite{Cromartie_2019}. The predicted mass-radius by the BigApple parameter set is consistent with the masses of the secondary component of GW190814 event \cite{RAbbott_2020}. The optimal lower bound on the maximum mass of non-rotating NS is recently determined with the help of the Universal relation connecting with mass and spins of uniformly rotating NS \cite{Most_2020}. The lower bound is reported $2.15\pm0.04 \ M_{\odot}$ by using the secondary mass of them GW190814, which is consistent with the IOPB-I parameter set (See table \ref{table4}). On the other hand, Tsokaros {\it et al.} \cite{Tsokaros_2020} claimed that the secondary object in GW190814 could be a slowly rotating NS with a relatively stiff EOS or a non-rotating NS with slightly stiff EOS. We also noticed that the representative BigApple EOS yields a radius of 12.96 km for the canonical star, which is consistent with NICER results. The NL3 model resides slightly above the bound of the secondary mass object of GW190814 due to its stiffer nature. 
\begin{figure} 
\centering
\includegraphics[width=0.6\textwidth]{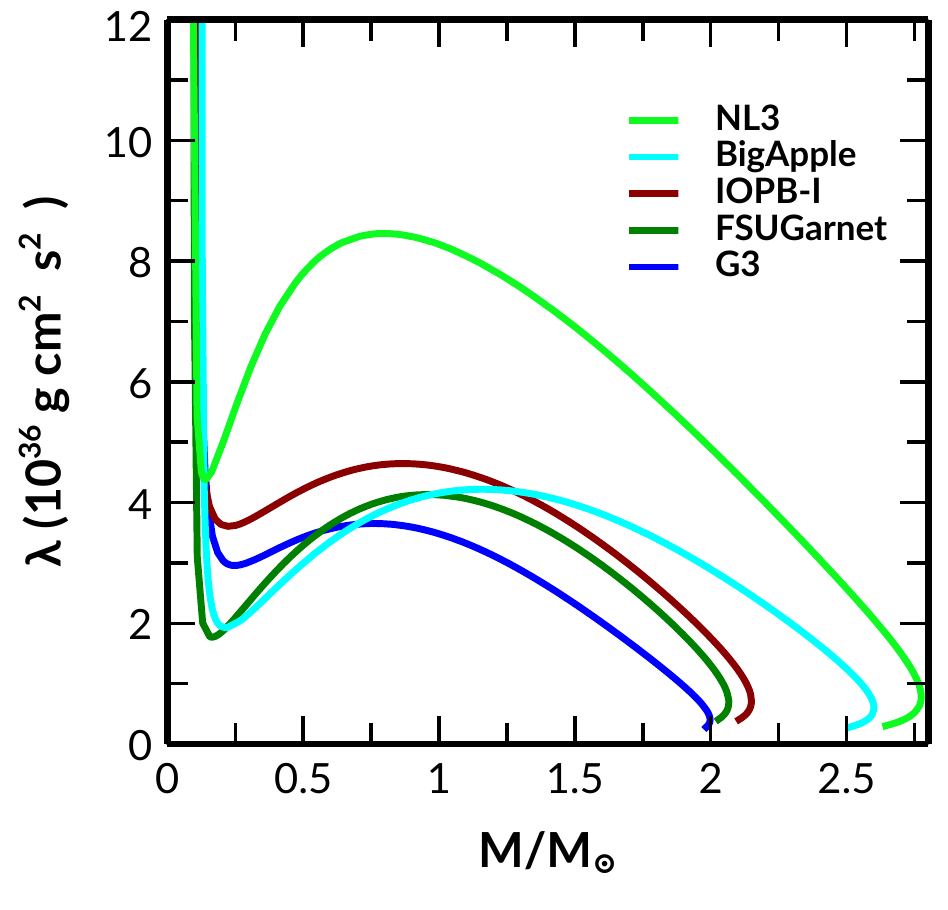}
\caption{(colour online) The tidal deformability $\lambda$ as a function of NS mass with different parameter sets.}
\label{lamb}
\end{figure}

Fig. \ref{lamb} represents the variation of $\lambda$ as a function of NS mass. NL3 model gives large tidal deformability and radius due to the stiff nature of the EOS at low and high densities. On the other hand, the BigApple parameter set shows EOS softer at high densities and stiffer at very low densities, yields a smaller radius and tidal deformability consistent with the observation. 

\begin{figure} 
\centering
\includegraphics[width=0.6\textwidth]{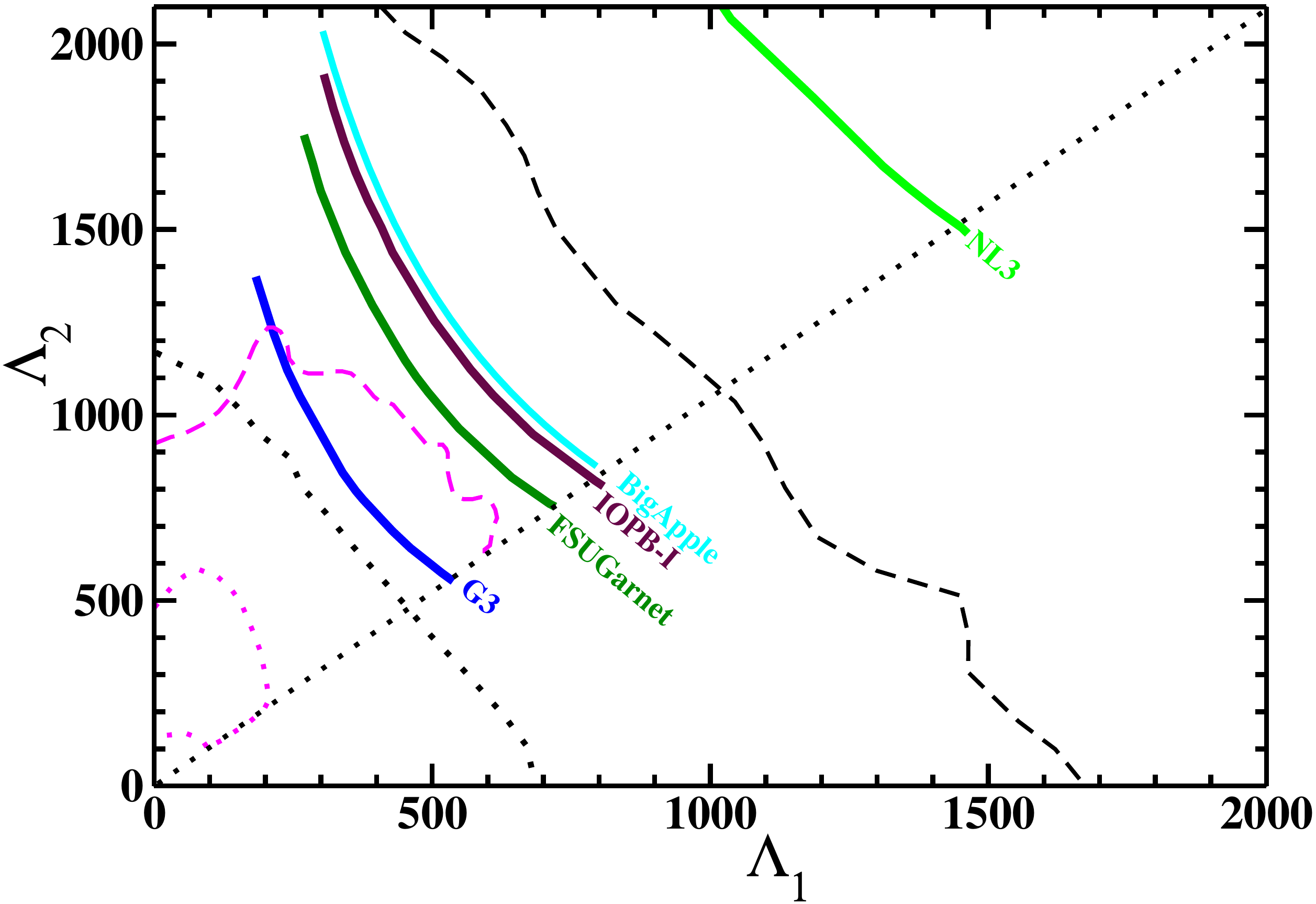}
\caption{(color online) Binary NS tidal deformabilities calculated for NL3, FSUGarnet, G3, IOPB-I and BigApple sets are compared with the $50\%$ (dotted) and $90\%$ (dashed) probability contour in case of low-spin scenario $|\chi| \leq 0.05$ as given by GW170817 \cite{Abbott_2017,Abbott_2018}.}
\label{tidal}
\end{figure}

In Fig. \ref{tidal}, we display the individual dimensionless tidal deformability $\Lambda_1$ and $\Lambda_2$ for a fixed chirp mass $1.188\ M_{\odot}$ which is associated with the binary NS merger event GW170817 \cite{Abbott_2017,Abbott_2018}. The tidal deformability predicted by five EOSs is calculated by varying the mass ratio ($q=M_2/M_1$) from 0.7 to 1.0, as inferred with the small NS spin. GW170817 disfavors the NL3 EOS lying outside the  50\% and 90\% probability contours.  We noticed that a smaller value of $\Lambda$  for fixed mass $M$ or softer EOS such as BigApple, G3, etc., satisfies the GW170817 observational data. The values of $\Lambda$ for five EOSs are given in Table \ref{table4}.

\begin{figure} 
\centering
\includegraphics[width=0.6\textwidth]{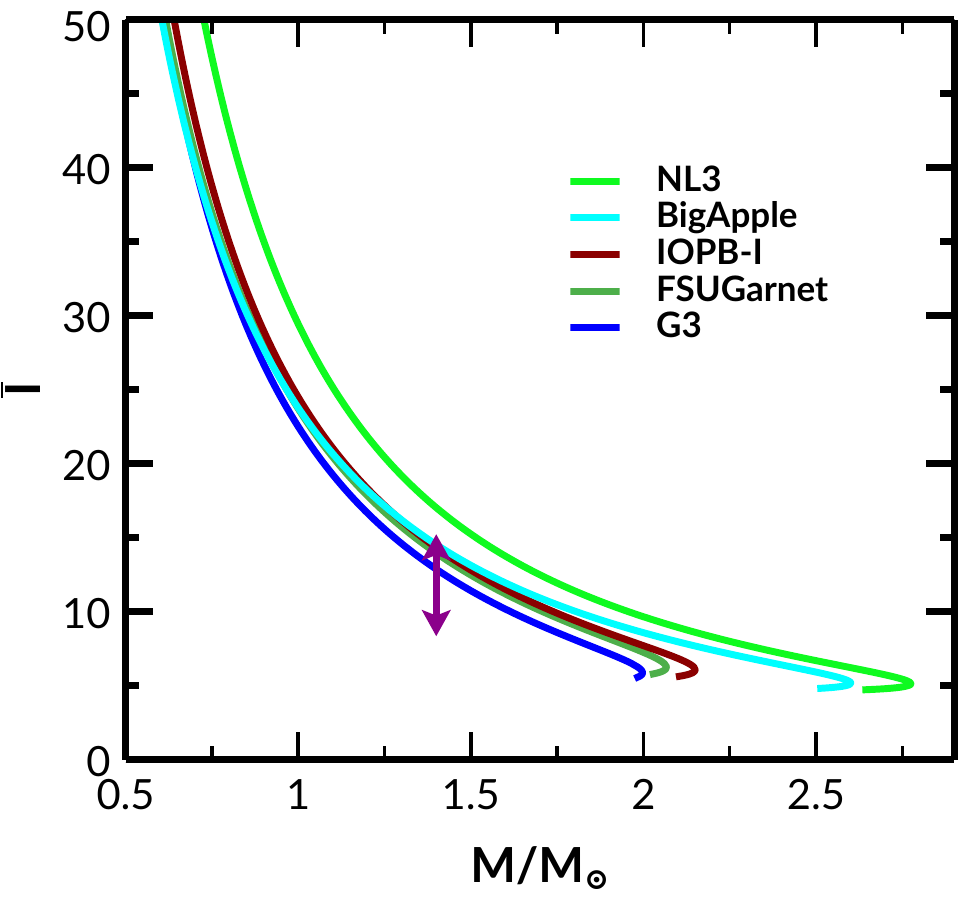}
\caption{(colour online) The dimensionless moment of inertia with the mass of the NS for different parameter sets. The overlaid arrows represent the MI constraint from the analysis of PSR J0737-3039A \cite{Landry_2018, Kumar_2019}.}
\label{mom}
\end{figure}

The moment of inertia (MI) of a slowly rotating NS (for a spherical star) is given by 
\begin{equation}
    I \approx \frac{8\pi}{3}\int_{0}^{R}({\cal E}+P)\ e^{-\phi(r)}\Big[1-\frac{2m(r)}{r}\Big]^{-1}\frac{\bar{\omega}}{\Omega}r^4 dr,
    \label{eqmom}
\end{equation}
where $\bar{\omega}$ and $\Omega$ are the dragging rotational functions and angular velocity respectively for the uniformly rotating NS \cite{Lattimer_2000, Krastev_2008}.  In Fig. \ref{mom}, we present the dimensionless MI ( $\bar{I}=I/M^3$) with the mass of the NS for five EOSs. The measurement of the MI of the pulsar PSR J0737+03039A has opened a new way to put a constraint on the nuclear EOSs \cite{Landry_2018, Kumar_2019}.  It may also be possible to refine some of the parameters that enter the NM models or even rule out all classes of models that currently exist. Fig. \ref{mom} shows that PSR 0737+03039A disfavors the MI corresponding to stiffer EOSs such as the NL3 set. However, the MI associated with softer EOS matches well with the deduced data. The calculated MI corresponds BigApple set just passes through the PSR data.

\section{Summary and Conclusions}
\label{summary}
The secondary component of the GW190814 event is either a supermassive NS or the lightest black hole because it has no tidal signatures and electromagnetic counterparts in the waveform. To explore this object a lot of debate exists whether it is the lightest black hole or heaviest neutron star. If it is the heaviest neutron star, the equation of state (EOS) must be stiffed. Therefore, we search for stiff EOS, which can reproduce the mass in that range. We find that there are only some density-dependent RMF models \cite{Huang_2020} which can reproduce the mass range given by the GW190814 event. Moreover, to explore this object, Fattoyev {\it et al.} have developed a new type of RMF parameter set (named as BigApple) which reproduces the mass $2.60 \ M_\odot$.

In this present paper, we systematically study the nuclear bulk properties such as binding energy per particle, charge radius, neutron-skin thickness, two-neutron separation energy, and single-particle energy, etc., for the magic nuclei Ca, Ni, Zr, Sn, Pb, and $Z=120$ in the whole isotopic chains. The predicted finite nuclei properties by the BigApple parameter set are compared with the NL3, IOPB-I, G3, and FSUGarnet set. It is found that all the finite nuclei properties are well predicted by BigApple and also reasonably consistent with the experimental data for the series of nuclei.

The NM properties such as incompressibility, symmetry energy, and slope parameter at the saturation density are almost satisfied by the BigApple set (see Table \ref{table2}). But the main theoretical uncertainty associated with BigApple is that it doesn't satisfy HIC data. This is because it is a stiff equation of state. Other constraints that don't respect by BigApple set are (i) Hebeler {\it et al.} data, (ii) symmetry energy from IAS, HIC Sn+Sn, and (iii)  ASY-EOS experimental data, etc. We compare the nuclear matter properties such as binding energy per particle and density-dependent symmetry energy for the BigApple case with different considered parameter sets. We find that the BigApple parameter set doesn't respect such constraints in the whole density limit. Therefore, we noticed that one can not use BigApple set to calculate the nuclear matter properties.

In the case of NS, the BigApple set predicted the NS's maximum mass and canonical radius as 2.60 $M_\odot$ and 12.96 km, respectively, which are consistent with GW190814 data. The canonical radius is also satisfied with NICER data. The calculated $\Lambda_{1.4}$ of the BigApple is 717.30, which is well suited with GW190814 data but disfavors the GW170817 data. Therefore, some observational properties of the NS predicted by the BigApple set are well consistent with GW190814. Therefore, it is advisable that one can use the BigApple set to calculate finite nuclei properties to some extent but not for nuclear matter and neutron stars.
\bibliography{BigApple}

\begin{thebibliography}{100}

\bibitem{Abbott_2017}
 The LIGO Scientific Collaboration and Virgo Collaboration Collaborations
  (B.~P. Abbott, R.~Abbott, T.~D. Abbott {\em et~al.}), {\em Phys. Rev. Lett.}
  {\bf 119} (Oct 2017)   161101.

\bibitem{Abbott_2018}
 The LIGO Scientific Collaboration and the Virgo Collaboration Collaborations
  (B.~P. Abbott, R.~Abbott, T.~D. Abbott {\em et~al.}), {\em Phys. Rev. Lett.}
  {\bf 121} (Oct 2018)   161101.

\bibitem{RAbbott_2020}
R.~Abbott, T.~D. Abbott, S.~Abraham {\em et~al.}, {\em The Astrophysical
  Journal} {\bf 896} (Jun 2020)   L44.

\bibitem{Most_2020}
E.~R. Most, L.~J. Papenfort, L.~R. Weih and L.~Rezzolla, {\em MNRAS} {\bf 499}
  (09 2020) L82.

\bibitem{Vattis_2020}
K.~Vattis, I.~S. Goldstein and S.~M. Koushiappas, {\em Phys. Rev. D} {\bf 102}
  (Sep 2020)   061301.

\bibitem{Tews_2021}
I.~Tews, P.~T.~H. Pang, T.~Dietrich, M.~W. Coughlin, S.~Antier, M.~Bulla,
  J.~Heinzel and L.~Issa, {\em The Astrophysical Journal} {\bf 908} (Feb 2021)
  ~L1.

\bibitem{Roupas_2021}
Z.~Roupas, {\em Astrophysics and Space Science} {\bf 366} (Jan 2021).

\bibitem{Godzieba_2021}
D.~A. Godzieba, D.~Radice and S.~Bernuzzi, {\em The Astrophysical Journal} {\bf
  908} (Feb 2021)   122.

\bibitem{Huang_2020}
K.~Huang, J.~Hu, Y.~Zhang and H.~Shen, {\em The Astrophysical Journal} {\bf
  904} (Nov 2020)  ~39.

\bibitem{Tan_2020}
H.~Tan, J.~Noronha-Hostler and N.~Yunes, {\em Phys. Rev. Lett.} {\bf 125} (Dec
  2020)   261104.

\bibitem{ZhangAAS_2020}
N.-B. Zhang and B.-A. Li, {\em The Astrophysical Journal} {\bf 902} (Oct 2020)
  ~38.

\bibitem{Fattoyev_2020}
F.~J. Fattoyev, C.~J. Horowitz, J.~Piekarewicz and B.~Reed, {\em Phys. Rev. C}
  {\bf 102} (Dec 2020)   065805.

\bibitem{DasPRD_2021}
H.~C. Das, A.~Kumar and S.~K. Patra, {\em Phys. Rev. D} {\bf 104} (Sep 2021)
  063028.

\bibitem{Biswas_2021}
B.~Biswas, R.~Nandi, P.~Char, S.~Bose and N.~Stergioulas, {\em Monthly Notices
  of the Royal Astronomical Society} {\bf 505} (05 2021) 1600.

\bibitem{Margalit_2017}
B.~Margalit and B.~D. Metzger, {\em The Astrophysical Journal} {\bf 850} (Nov
  2017)   L19.

\bibitem{Rezzolla_2018}
L.~Rezzolla, E.~R. Most and L.~R. Weih, {\em The Astrophysical Journal} {\bf
  852} (Jan 2018)   L25.

\bibitem{Shibata_2019}
M.~Shibata, E.~Zhou, K.~Kiuchi and S.~Fujibayashi, {\em Phys. Rev. D} {\bf 100}
  (Jul 2019)   023015.

\bibitem{Demorest_2010}
P.~B. Demorest, T.~Pennucci, S.~M. Ransom, M.~S.~E. Roberts and J.~W.~T.
  Hessels, {\em Nature} {\bf 467} (Oct 2010)   1081–1083.

\bibitem{Antoniadis_2013}
J.~Antoniadis, P.~C.~C. Freire, N.~Wex {\em et~al.}, {\em Science} {\bf 340}
  (2013)   1233232.

\bibitem{Cromartie_2019}
H.~T. Cromartie, E.~Fonseca, S.~M. Ransom, P.~B. Demorest {\em et~al.}, {\em
  Nature Astronomy} {\bf 4} (Sep 2019)   72–76.

\bibitem{Miller_2019}
M.~C. Miller, F.~K. Lamb, A.~J. Dittmann {\em et~al.}, {\em APJ} {\bf 887} (Dec
  2019)   L24.

\bibitem{Riley_2019}
T.~E. Riley, A.~L. Watts, S.~Bogdanov {\em et~al.}, {\em APJ} {\bf 887} (Dec
  2019)   L21.

\bibitem{Raaijmakers_2019}
G.~Raaijmakers, T.~E. Riley, A.~L. Watts {\em et~al.}, {\em Astrophys. J} {\bf
  887} (Dec 2019)   L22.

\bibitem{Lalazissis_1997}
G.~A. Lalazissis, J.~K\"onig and P.~Ring, {\em Phys. Rev. C} {\bf 55} (Jan
  1997) 540.

\bibitem{Frun_1997}
R.~J. Furnstahl, B.~D. Serot and H.-B. Tang, {\em Nucl. Phys. A} {\bf 615}
  (1997) 441 .

\bibitem{Reinhard_1988}
P.~G. Reinhard, {\em Z. Phys. A Atomic Nuclei} {\bf 329} (Sep 1988) 257.

\bibitem{Singh_2013}
S.~K. Singh, M.~Bhuyan, P.~K. Panda and S.~K. Patra, {\em Journal of Physics G:
  Nuclear and Particle Physics} {\bf 40} (Jul 2013)   085104.

\bibitem{Kumar_2017}
B.~Kumar, S.~Singh, B.~Agrawal and S.~Patra, {\em Nuclear Physics A} {\bf 966}
  (2017) 197 .

\bibitem{Kumar_2018}
B.~Kumar, S.~K. Patra and B.~K. Agrawal, {\em Phys. Rev. C} {\bf 97} (Apr 2018)
    045806.

\bibitem{Chabanta_1998}
E.~Chabanat, P.~Bonche, P.~Haensel, J.~Meyer and R.~Schaeffer", {\em Nucl.
  Phys. A} {\bf 635}  (1998) 231 .

\bibitem{Brown_1998}
B.~Alex~Brown, {\em Phys. Rev. C} {\bf 58} (Jul 1998) 220.

\bibitem{Stone_2007}
J.~Stone and P.-G. Reinhard, {\em Progress in Particle and Nuclear Physics}
  {\bf 58} (Apr 2007)   587–657.

\bibitem{Dutra_2012}
M.~Dutra, O.~Louren\ifmmode~\mbox{\c{c}}\else \c{c}\fi{}o, J.~S. S\'a~Martins
  {\em et~al.}, {\em Phys. Rev. C} {\bf 85} (Mar 2012)   035201.

\bibitem{Niksic_2002}
T.~Nik\ifmmode \check{s}\else \v{s}\fi{}i\ifmmode~\acute{c}\else \'{c}\fi{},
  D.~Vretenar, P.~Finelli and P.~Ring, {\em Phys. Rev. C} {\bf 66} (Aug 2002)
  024306.

\bibitem{Typel_2005}
S.~Typel, {\em Phys. Rev. C} {\bf 71} (Jun 2005)   064301.

\bibitem{Lalazissis_2005}
G.~A. Lalazissis, T.~Nik\ifmmode \check{s}\else
  \v{s}\fi{}i\ifmmode~\acute{c}\else \'{c}\fi{}, D.~Vretenar and P.~Ring, {\em
  Phys. Rev. C} {\bf 71} (Feb 2005)   024312.

\bibitem{Klahn_2006}
T.~Kl\"ahn, D.~Blaschke, S.~Typel {\em et~al.}, {\em Phys. Rev. C} {\bf 74}
  (Sep 2006)   035802.

\bibitem{Dutra_2014}
M.~Dutra, O.~Louren\ifmmode~\mbox{\c{c}}\else \c{c}\fi{}o, S.~S. Avancini {\em
  et~al.}, {\em Phys. Rev. C} {\bf 90} (Nov 2014)   055203.

\bibitem{Danielewicz_2002}
P.~Danielewicz, R.~Lacey and W.~G. Lynch, {\em Science} {\bf 298}  (2002) 1592.

\bibitem{Bauswein_2017}
A.~Bauswein, O.~Just, H.-T. Janka and N.~Stergioulas, {\em The Astrophysical
  Journal} {\bf 850} (Nov 2017)   L34.

\bibitem{Annala_2018}
E.~Annala, T.~Gorda, A.~Kurkela and A.~Vuorinen, {\em Phys. Rev. Lett.} {\bf
  120} (Apr 2018)   172703.

\bibitem{Fattoyev_2018}
F.~J. Fattoyev, J.~Piekarewicz and C.~J. Horowitz, {\em Phys. Rev. Lett.} {\bf
  120} (Apr 2018)   172702.

\bibitem{Radice_2018}
D.~Radice, A.~Perego, F.~Zappa and S.~Bernuzzi, {\em The Astrophysical Journal}
  {\bf 852} (Jan 2018)   L29.

\bibitem{Mallik_2018}
T.~Malik, N.~Alam, M.~Fortin and Fothers, {\em Phys. Rev. C} {\bf 98} (Sep
  2018)   035804.

\bibitem{Most_2018}
E.~R. Most, L.~R. Weih, L.~Rezzolla and J.~Schaffner-Bielich, {\em Phys. Rev.
  Lett.} {\bf 120} (Jun 2018)   261103.

\bibitem{Tews_2018}
I.~Tews, J.~Margueron and S.~Reddy, {\em Phys. Rev. C} {\bf 98} (Oct 2018)
  045804.

\bibitem{Nandi_2019}
R.~Nandi, P.~Char and S.~Pal, {\em Phys. Rev. C} {\bf 99} (May 2019)   052802.

\bibitem{Capano_2020}
C.~D. Capano, I.~Tews, S.~M. Brown {\em et~al.}, {\em Nature Astronomy} {\bf 4}
  (Mar 2020)   625–632.

\bibitem{Bogdanov_2019}
S.~Bogdanov, F.~K. Lamb, S.~Mahmoodifar {\em et~al.}, {\em APJ} {\bf 887} (Dec
  2019)   L26.

\bibitem{Singh_2014}
S.~K. Singh, S.~K. Biswal, M.~Bhuyan and S.~K. Patra, {\em Journal of Physics
  G: Nuclear and Particle Physics} {\bf 41} (Mar 2014)   055201.

\bibitem{Chen_2015}
W.-C. Chen and J.~Piekarewicz, {\em Physics Letters B} {\bf 748}  (2015) 284 .

\bibitem{Negele_1970}
J.~W. Negele, {\em Phys. Rev. C} {\bf 1} (Apr 1970) 1260.

\bibitem{MCentelles_2001}
M.~Del~Estal, M.~Centelles, X.~Vi\~nas and S.~K. Patra, {\em Phys. Rev. C} {\bf
  63} (Jan 2001)   024314.

\bibitem{DelEstal_2001}
M.~Del~Estal, M.~Centelles, X.~Vi\~nas and S.~K. Patra, {\em Phys. Rev. C} {\bf
  63} (Mar 2001)   044321.

\bibitem{Das_2020}
H.~C. Das, A.~Kumar, B.~Kumar {\em et~al.}, {\em MNRAS} {\bf 495} (05 2020)
  4893.

\bibitem{Horowitz_2014}
C.~J. Horowitz, E.~F. Brown, Y.~Kim {\em et~al.}, {\em Journal of Physics G:
  Nuclear and Particle Physics} {\bf 41} (Jul 2014)   093001.

\bibitem{Baldo_2016}
M.~Baldo and G.~Burgio, {\em Progress in Particle and Nuclear Physics} {\bf 91}
  (Nov 2016)   203–258.

\bibitem{BaoLi_2013}
B.-A. Li and X.~Han, {\em Physics Letters B} {\bf 727}  (2013) 276 .

\bibitem{BaoLi_2019}
B.-A. Li, P.~G. Krastev, D.-H. Wen and N.-B. Zhang, {\em EPJ A} {\bf 55} (Jul
  2019).

\bibitem{Matsui_1981}
T.~Matsui, {\em Nucl. Phys. A} {\bf 370}  (1981) 365 .

\bibitem{Kubis_1997}
S.~Kubis and M.~Kutschera, {\em Phys. Lett. B} {\bf 399} (May 1997)
  191–195.

\bibitem{Chen_2014}
W.-C. Chen and J.~Piekarewicz, {\em Phys. Rev. C} {\bf 90} (Oct 2014)   044305.

\bibitem{Bethe_1971}
H.~A. Bethe, {\em Annual Review of Nuclear Science} {\bf 21}  (1971) 93.

\bibitem{Danielewicz_2014}
P.~Danielewicz and J.~Lee, {\em Nucl. Phys. A} {\bf 922}  (2014) 1 .

\bibitem{Zimmerman_2020}
J.~Zimmerman, Z.~Carson, K.~Schumacher, A.~W. Steiner and K.~Yagi,  (2020).

\bibitem{Colo_2014}
G.~Col{\`o}, U.~Garg and H.~Sagawa, {\em EPJ A} {\bf 50} (Feb 2014)  ~26.

\bibitem{Stone_2014}
J.~R. Stone, N.~J. Stone and S.~A. Moszkowski, {\em Phys. Rev. C} {\bf 89} (Apr
  2014)   044316.

\bibitem{Pearson_2010}
J.~M. Pearson, N.~Chamel and S.~Goriely, {\em Phys. Rev. C} {\bf 82} (Sep 2010)
    037301.

\bibitem{TLi_2010}
T.~Li, U.~Garg, Y.~Liu {\em et~al.}, {\em Phys. Rev. C} {\bf 81} (Mar 2010)
  034309.

\bibitem{Wang_2012}
M.~Wang, G.~Audi, A.~Wapstra {\em et~al.}, {\em Chinese Physics C} {\bf 36}
  (Dec 2012) 1603.

\bibitem{Angeli_2013}
I.~Angeli and K.~Marinova, {\em Atomic Data and Nuclear Data Tables} {\bf 99}
  (2013) 69 .

\bibitem{Abrahamyan_2012}
 PREX Collaboration Collaboration (S.~Abrahamyan {\em et~al.}), {\em Phys. Rev.
  Lett.} {\bf 108} (Mar 2012)   112502.

\bibitem{Adhikari_2021}
 PREX Collaboration Collaboration (D.~Adhikari, H.~Albataineh, D.~Androic {\em
  et~al.}), {\em Phys. Rev. Lett.} {\bf 126} (Apr 2021)   172502.

\bibitem{Reed_2021}
B.~T. Reed, F.~J. Fattoyev, C.~J. Horowitz and J.~Piekarewicz, {\em Phys. Rev.
  Lett.} {\bf 126} (Apr 2021)   172503.

\bibitem{Pattnaik_2021}
J.~A. Pattnaik, R.~N. Panda, M.~Bhuyan and S.~K. Patra,  (2021).

\bibitem{Trzci_2001}
A.~Trzci\ifmmode~\acute{n}\else \'{n}\fi{}ska,
  J.~Jastrz\ifmmode~\mbox{\c{e}}\else \c{e}\fi{}bski,
  P.~Lubi\ifmmode~\acute{n}\else \'{n}\fi{}ski {\em et~al.}, {\em Phys. Rev.
  Lett.} {\bf 87} (Aug 2001)   082501.

\bibitem{Zenihiro_2010}
J.~Zenihiro, H.~Sakaguchi, T.~Murakami {\em et~al.}, {\em Phys. Rev. C} {\bf
  82} (Oct 2010)   044611.

\bibitem{Vautherin_1972}
D.~Vautherin and D.~M. Brink, {\em Phys. Rev. C} {\bf 5} (Mar 1972) 626.

\bibitem{Moller_2016}
P.~Möller, A.~Sierk, T.~Ichikawa and H.~Sagawa, {\em Atomic Data and Nuclear
  Data Tables} {\bf 109-110}  (2016) 1 .

\bibitem{Rutz_1997}
K.~Rutz, M.~Bender, T.~B\"urvenich {\em et~al.}, {\em Phys. Rev. C} {\bf 56}
  (Jul 1997) 238.

\bibitem{Gupta_1997}
R.~K. Gupta, S.~K. Patra and W.~Greiner, {\em Modern Physics Letters A} {\bf
  12}  (1997) 1727.

\bibitem{Patra_1999}
S.~K. Patra, C.-L. Wu, C.~R. Praharaj and R.~K. Gupta, {\em Nuclear Physics A}
  {\bf 651}  (1999) 117 .

\bibitem{Mehta_2015}
M.~S. Mehta, H.~Kaur, B.~Kumar and S.~K. Patra, {\em Phys. Rev. C} {\bf 92}
  (Nov 2015)   054305.

\bibitem{Bhuyan_2012}
M.~Bhuyan and S.~K. Patra, {\em Mod. Phys. Lett. A} {\bf 27}  (2012)   1250173.

\bibitem{Sharma_1995}
M.~M. Sharma, G.~Lalazissis, J.~K\"onig and P.~Ring, {\em Phys. Rev. Lett.}
  {\bf 74} (May 1995) 3744.

\bibitem{Hebeler_2013}
K.~Hebeler, J.~M. Lattimer, C.~J. Pethick and A.~Schwenk, {\em The
  Astrophysical Journal} {\bf 773} (Jul 2013)  ~11.

\bibitem{Gezerlis_2010}
A.~Gezerlis and J.~Carlson, {\em Phys. Rev. C} {\bf 81} (Feb 2010)   025803.

\bibitem{Baldo_2008}
M.~Baldo and C.~Maieron, {\em Phys. Rev. C} {\bf 77} (Jan 2008)   015801.

\bibitem{Friedman_1981}
B.~Friedman and V.~Pandharipande, {\em Nuclear Physics A} {\bf 361}  (1981) 502
  .

\bibitem{Gandolfi_2008}
S.~Gandolfi, A.~Y. Illarionov, S.~Fantoni, F.~Pederiva and K.~E. Schmidt, {\em
  Phys. Rev. Lett.} {\bf 101} (Sep 2008)   132501.

\bibitem{Tsang_2009}
M.~B. Tsang, Y.~Zhang, P.~Danielewicz {\em et~al.}, {\em Phys. Rev. Lett.} {\bf
  102} (Mar 2009)   122701.

\bibitem{Tsang_2010}
M.~B. Tsang {\em et~al.}, {\em IJMPE} {\bf 19}  (2010) 1631.

\bibitem{Russotto_2016}
P.~Russotto {\em et~al.}, {\em Phys. Rev. C} {\bf 94} (Sep 2016)   034608.

\bibitem{Dorso_2019}
C.~Dorso, G.~Frank and J.~López, {\em Nuclear Physics A} {\bf 984}  (2019) 77
  .

\bibitem{Zhang_2020}
Y.~Zhang, M.~Liu, C.-J. Xia, Z.~Li and S.~K. Biswal, {\em Phys. Rev. C} {\bf
  101} (Mar 2020)   034303.

\bibitem{Lattimer_2014}
J.~M. Lattimer, {\em Nuclear Physics A} {\bf 928}  (2014) 276 .

\bibitem{Gandolfi_2016}
S.~Gandolfi and A.~W. Steiner, {\em Journal of Physics: Conference Series} {\bf
  665} (Jan 2016)   012063.

\bibitem{BKS_2015}
B.~K. Sharma, M.~Centelles, X.~Vi\~nas, M.~Baldo and G.~F. Burgio, {\em A\&A}
  {\bf 584} (Nov 2015)   A103.

\bibitem{TOV1}
R.~C. {Tolman}, {\em Phys. Rev.} {\bf 55} (Feb 1939) 364.

\bibitem{TOV2}
J.~R. Oppenheimer and G.~M. Volkoff, {\em Phys. Rev.} {\bf 55} (Feb 1939) 374.

\bibitem{Hinderer_2010}
T.~Hinderer, B.~D. Lackey, R.~N. Lang and J.~S. Read, {\em Phys. Rev. D} {\bf
  81} (Jun 2010)   123016.

\bibitem{Kumartidal_2017}
B.~Kumar, S.~K. Biswal and S.~K. Patra, {\em Phys. Rev. C} {\bf 95} (Jan 2017)
   015801.

\bibitem{Hinderer_2008}
T.~Hinderer, {\em The Astrophysical Journal} {\bf 677} (Apr 2008) 1216.

\bibitem{Tsokaros_2020}
A.~Tsokaros, M.~Ruiz and S.~L. Shapiro, {\em The Astrophysical Journal} {\bf
  905} (Dec 2020)  ~48.

\bibitem{Lattimer_2000}
J.~M. Lattimer and M.~Prakash, {\em Physics Reports} {\bf 333-334} (Aug 2000)
  121–146.

\bibitem{Krastev_2008}
P.~G. Krastev, B.~Li and A.~Worley, {\em APJ} {\bf 676} (Apr 2008)
  1170–1177.

\bibitem{Landry_2018}
P.~Landry and B.~Kumar, {\em APJ} {\bf 868} (Nov 2018)   L22.

\bibitem{Kumar_2019}
B.~Kumar and P.~Landry, {\em Phys. Rev. D} {\bf 99} (Jun 2019)   123026.

\end{thebibliography}
\bibliographystyle{ws-ijmpe}
\end{document}